%% file: main.tex
\documentclass[aps,pra,reprint]{revtex4-2}
\usepackage[utf8]{inputenc}
\usepackage{graphicx}
\usepackage{amsmath}
\usepackage{comment}
\usepackage{siunitx}
\usepackage[caption=false]{subfig}
\usepackage{subfiles}
\usepackage{quantikz}
\usepackage{bm}%
\usepackage{array}

\usepackage[colorlinks]{hyperref}
\hypersetup{%
        plainpages=true,
        breaklinks=true,
        hypertexnames=false,
        pageanchor=true,
        colorlinks=true,
        linkcolor={blue},
        citecolor={blue},
        urlcolor={blue},
        anchorcolor={black}
      }

\renewcommand{\eqref}[1]{\mbox{Eq.~(\ref{#1})}}
\newcommand{\tabref}[1]{\mbox{Table~\ref{#1}}}
\newcommand{\figref}[1]{\mbox{Fig.~\ref{#1}}}
\newcommand{\figpanel}[2]{Fig.~\hyperref[#1]{\ref*{#1}(#2)}} 
\newcommand{\figpanels}[3]{Fig.~\hyperref[#1]{\ref*{#1}(#2)-(#3)}} 
\newcommand{\figpanelNoPrefix}[2]{\hyperref[#1]{\ref*{#1}(#2)}} 
\newcommand{\figpanelsNoPrefix}[3]{\hyperref[#1]{\ref*{#1}(#2)-(#3)}} 

\usepackage{mleftright} 

\begin{document}

\title{Lattice surgery with Bell measurements: Modular fault-tolerant quantum computation at low entanglement cost}
\author{Trond Hjerpekjøn Haug}
\email{trond.haug@chalmers.se}
\author{Timo Hillmann}
\author{Anton Frisk Kockum}
\author{Rapha\"{e}l Van Laer}
\email{raphael.van.laer@chalmers.se}
\affiliation{Department of Microtechnology and Nanoscience, Chalmers University of Technology, 412 96 Gothenburg, Sweden}
\date{\today}
\begin{abstract}
    Modular architectures are a promising approach to scaling quantum computers to fault tolerance. 
    Small, low-noise quantum processors connected through relatively noisy quantum links are capable of fault-tolerant operation as long as the noise can be confined to the interface. 
    Finding protocols that implement the quantum links between modules as efficiently as possible is essential because inter-module entanglement is challenging to produce at a similar rate and fidelity as local entanglement. 
    We introduce a protocol for lattice surgery on surface codes in which all non-local operations are Bell measurements. 
    The protocol simultaneously confines the link noise and requires only half as many module-crossing gates as previously proposed protocols.
    To mitigate distance-reducing hook errors, we introduce a strategy of alternating the gate sequence between rounds of syndrome measurement, which prevents multiple hooks from simultaneously aligning with a logical operator in the code. 
    We evaluate our protocol's performance when two logical qubits on separate modules are prepared in a logical Bell state. 
    Circuit-level simulations under depolarizing noise show that the logical error suppression for a given entanglement rate between modules is consistently stronger compared to the best-performing alternative protocols for a wide range of link noise, with a typical $\sim\SI{40}{\percent}$ entanglement resource saving for a constant logical error rate. 
    Our approach to protocol design is applicable to any quantum circuit that must be divided across processor modules and can therefore guide development of resource-efficient modular quantum computation beyond the surface code. 
    
\end{abstract}
\maketitle

\section{Introduction}
Quantum computation and communication offer new opportunities that are believed to go beyond what can be achieved classically~\cite{shor_polynomial-time_1997, Gisin2007, Dalzell2023, huang_vast_2025}.
However, to unlock these capabilities, protecting quantum information from the noise of physical hardware components is necessary. 
This need has motivated the development of quantum error correction and fault tolerance, which introduce redundancy in the storage and manipulation of quantum information to protect it against noise. 

Scaling up the number of physical components to meet the demands of fault tolerance creates challenges for hardware platforms currently in development. 
For example, superconducting qubits~\cite{Gu2017, Blais2021} are currently cryogenically cooled and controlled using cables connected to room-temperature electronics. 
Increasing the number of qubits, therefore, increases the demand for cooling power from cryostats, which often is in short supply~\cite{krinner_engineering_2019, mohseni_how_2025}. 
Moreover, fabrication imperfections and crosstalk limit the precision of qubit operations, and the probability of critical defects increases rapidly with system size~\cite{kosen_signal_2024}. 
Other platforms, e.g., neutral atoms~\cite{Strohm2024} and trapped ions~\cite{Wintersperger2023}, do not suffer in the same way from fabrication variations but are similarly susceptible to crosstalk.
Furthermore, the requirement for chambers with ultra-high vacuum conditions ($\sim 10^{11}$ torr) to accommodate the qubits in these platforms poses a significant obstacle to scalability~\cite{monroe_scaling_2013}.

Modular architectures have been proposed as a solution to overcome the challenges faced by monolithic systems~\cite{jiang_distributed_2007, monroe_large-scale_2014, nickerson_topological_2013, nickerson_freely_2014, li_hierarchical_2016, aghaee_rad_scaling_2025}. 
In such architectures, small processors with low levels of noise can be connected through quantum links that allow them to operate as a single entity in a network configuration. 
Depending on the size of each module, it has been found that the noise suffered by the links can be about an order of magnitude higher than local levels without significantly compromising the logical error suppression of surface codes~\cite{ramette_fault-tolerant_2024}.  
The resilience to noise at the interface suggests that focus can partly shift from solving the scaling challenges in monolithic architectures to establishing relatively error-prone quantum links between modules.
To prevent bottlenecks, the bandwidth of these inter-module quantum links should be comparable to the internal clock rate of the quantum processor modules. 
Links can be short-range interconnects that support direct gates between modules~\cite{gold_entanglement_2021, wu_modular_2024, mollenhauer_high-efficiency_2024} or long-range connections for distribution of entanglement using itinerant photons~\cite{roch_observation_2014, hensen_loophole-free_2015, dickel_chip_2018, zhong_deterministic_2021,burkhart_error_2021, main_distributed_2025, almanakly_deterministic_2025}.

\begin{figure*}
    \centering
    \includegraphics[width=\linewidth]{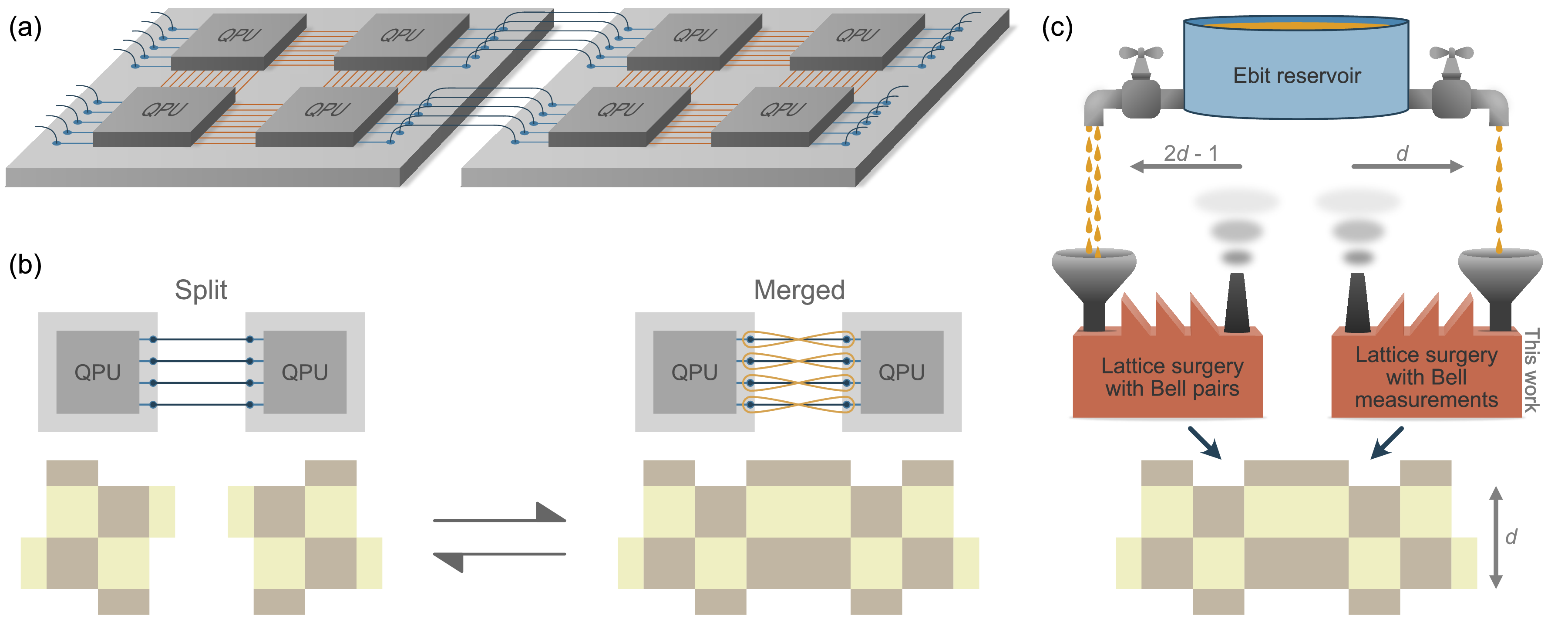}
    \caption{Modular quantum-computing architectures and lattice surgery on surface codes. 
    (a) An example modular architecture for a superconducting quantum computer with four quantum processing units (QPUs) connected by direct links (orange) through an interposer chip. 
    Each QPU is also equipped with photonic connections (blue) that support entanglement distribution with itinerant photons to a QPU on another module. 
    (b) Lattice surgery with surface-code logical qubits on separate modules connected by quantum links. 
    Dark (light) squares represent X(Z)-stabilizers of the surface code.
    When patches are merged, physical qubits on separate modules interact by consuming entanglement that must be established between modules over the quantum links. 
    (c) The rate of entanglement consumption depends on the protocol that is used for the surgery. 
    Previous protocols have required $2d-1$ Bell pairs per round of stabilizer measurement in a code of distance $d$. Each Bell pair has one ebit of entanglement~\cite{bennett_concentrating_1996}.
    We introduce a protocol for lattice surgery based on Bell measurements, which only requires $d$ ebits per round of syndrome measurement.
    }
    \label{fig:chip}
\end{figure*}

In this work, we consider a modular architecture of quantum processing units (QPUs) where each unit supports at least one surface-code logical qubit.
The QPUs are connected through quantum links that enable either direct gates or the distribution of Bell pairs between modules. 
An example architecture is shown in \figpanel{fig:chip}{a}. 
Operations on logical qubits between separate modules are performed using lattice surgery~\cite{horsman_surface_2012, litinski_game_2019}---a popular method for executing logical operations on topological codes like the surface code~\cite{kitaev_fault-tolerant_2003, dennis_topological_2002}.

We define a ``patch'' as the group of physical qubits that collectively support one surface-code logical qubit. 
By merging two or more patches at their boundaries, we can measure a product of logical operators that are encoded within the patches.
This merging of patches entangles the physical qubits from one patch with those of the other patch [\figpanel{fig:chip}{b}]. 
While the patches are merged, products of Pauli operators on each patch can be inferred from the measurement outcomes of the new stabilizers with support on both patches. 
After a product of logical operators has been measured fault-tolerantly, the merged patches are split so that logical qubits are free to take part in the next step of the algorithm. 
Splitting is performed by measuring a subset of the physical qubits in the merged patch.

Recent studies of rotated surface codes in modular architectures have introduced protocols for merging patches across an interface~\cite{shalby_optimized_2025,jacinto_network_2025}. 
A protocol called ``cat state gadget" in Ref.~\cite{shalby_optimized_2025} and ``measurement teleportation" in Ref.~\cite{jacinto_network_2025} uses Bell pairs to measure stabilizers at the interface. 
It can be implemented with only $d$ Bell pairs per round of syndrome measurements, where $d$ is the distance of the surface code, at the cost of cutting the effective code distance at the interface in half.
This distance reduction, as well as the noise propagation from the Bell pairs to data qubits, implies that the protocol is impractical.  
Full distance in the protocol can be restored by increasing the number of Bell pairs per round to $2d-1$. 
Distribution of high-fidelity Bell pairs between processor modules is currently a difficult and slow process~\cite{magnard_microwave_2020, stephenson_high-rate_2020, heya_randomized_2025,almanakly_deterministic_2025}, which suggests that Bell pairs are a valuable resource that should be used sparingly.
However, to our knowledge, there is no protocol for inter-module lattice surgery with an effective distance at the interface larger than $(d+1)/2$ which requires less than $2d-1$ Bell pairs per round of syndrome measurement. 

Here, we develop such a protocol based on Bell measurements [\figpanel{fig:chip}{c}].
A Bell measurement is a projective measurement of two qubits onto a Bell state. 
It measures the observables $X_1 X_2$ and $Z_1 Z_2$ simultaneously, where $X_i$ and $Z_i$ are Pauli operators on the $i^{\mathrm{th}}$ qubit. 
We refer the reader to Appendix~\ref{sec:Bell_circuits} for details on Bell measurements and their implementation with quantum circuits.  
Bell measurements are ubiquitous in discrete-value optical quantum technologies, where they are often referred to as fusion gates~\cite{browne_resource-efficient_2005}. 
The fusion-based model of quantum computation~\cite{bartolucci_fusion_2023} employs Bell measurements on entangled resource states.
The key insight is that Bell measurements---like multibody entangling measurements on Bell pairs~\cite{verstraete_valence_2004}---create the entanglement that is required by the code, as well as the syndrome information to detect errors~\cite{walshe_linear-optical_2025}.  
We use this insight to construct a gate-based protocol for lattice surgery between modules in which Bell measurements replace the Bell pairs used in Refs.~\cite{shalby_optimized_2025, jacinto_network_2025}.
We also introduce a technique for controlling distance-reducing errors at the interface by alternating the syndrome-measurement circuit between rounds of syndrome measurement.
Finally, using a realistic circuit-level noise model, we compare the performance of the Bell-measurement protocol to a benchmark protocol from Ref.~\cite{shalby_optimized_2025}.
For physically reasonable parameters, we find that using Bell measurements reduces the required entanglement rate between modules by $\sim\SI{40}{\percent}$ compared to the benchmark protocol for a constant logical error rate. 
Our results demonstrate that the Bell-measurement protocol achieves a stronger logical error suppression than previously proposed protocols when access to high-fidelity inter-module entanglement is limited.

This paper is organized as follows. 
In Section~\ref{sec:Bell_measurements}, we introduce the benchmark and Bell-measurement protocols using the language of ZX calculus. 
We explain in detail the effects of using Bell measurements for lattice surgery on rotated surface codes in a circuit-based implementation. 
In Section~\ref{sec:circuit_wiggle}, we address the problem of distance reduction at the interface, which requires an alternating syndrome-measurement circuit to mitigate. Numerical studies of the performance of the Bell-measurement protocol are presented in  Section~\ref{sec:performance} and compared to the benchmark protocol under the same levels of noise. We conclude and present an outlook for the Bell-measurement protocol in Section~\ref{sec:conclusion_outlook}.

\section{Protocol design and effects of Bell measurements}
\label{sec:Bell_measurements}
A well-designed protocol for inter-module lattice surgery should have the following properties. 
(i) The protocol should ensure that noise from operations at the interface remains confined to the interface. 
Two-qubit gates applied locally can propagate errors that originate at the noisy interface to the bulk. 
This complicates error tracking and could lead to a reduction in the effective code distance at the interface.  
(ii) Operations at the interface should not interfere with operations in the bulk.
For example, if the gate sequence at the interface for a round of syndrome measurement is significantly longer than the corresponding sequence in the bulk, the added circuit depth slows down the logical speed of the processor and allows more noise to accumulate before each syndrome measurement.  
(iii) If direct gates between physical qubits are not supported by the architecture and access to high-fidelity Bell pairs is limited, the protocol should involve as few teleported gates as possible at each round of syndrome measurement. 
Each teleported gate consumes one Bell pair that must be established between modules before the gate can be applied. 

One possible protocol is to keep the same gate sequence at the interface as in the bulk. 
Each weight-4 stabilizer at the interface is measured using four CX or CZ gates between the data qubits and a single ancilla qubit.
The ancilla qubit can be in either of the two modules. 
Since two data qubits will be in the module without the ancilla qubit, two of the four gates must be applied across an interface and will therefore introduce elevated noise. 
This protocol avoids distance-reducing errors, also known as hook errors, and does not interfere with the gate schedule in the bulk. 
However, the noisy interface-crossing gates will increase the error rate on data qubits at the interface. 
Such errors can subsequently propagate through local two-qubit gates. 
If the architecture does not allow the interface-crossing gates to be implemented directly, the protocol needs $2d-1$ Bell pairs per round of syndrome measurement to teleport the gates. 
We refer to this protocol as the benchmark protocol.
It was found to be the protocol that achieved the lowest logical error rate of those studied in Ref.~\cite{shalby_optimized_2025}.

A protocol for lattice surgery based on Bell measurements cannot easily be understood in terms of a code at a single instance in time.  
It is more insightful to take a holistic view of the structure that emerges as the code evolves through time~\cite{hillmann_single-shot_2024}, also referred to as the code in space-time.
In this picture, outcomes from Bell measurements at different points in space and time are combined to form detectors. 
A detector is a set of binary measurement outcomes $D = \{ m_i\}$ whose parity is deterministic in the absence of errors~\cite{mcewen_relaxing_2023}.
An error that anticommutes with a measurement used to form a detector will flip the detector's parity.
In the surface code, a single-qubit $X$ or $Z$ error flips the parity of at most two detectors. A $Y$ error is equivalent to a combined $ X$- and $Z$ error and thus flips the parity of four detectors. 

\begin{figure*}
    \centering
    \includegraphics[width=\linewidth]{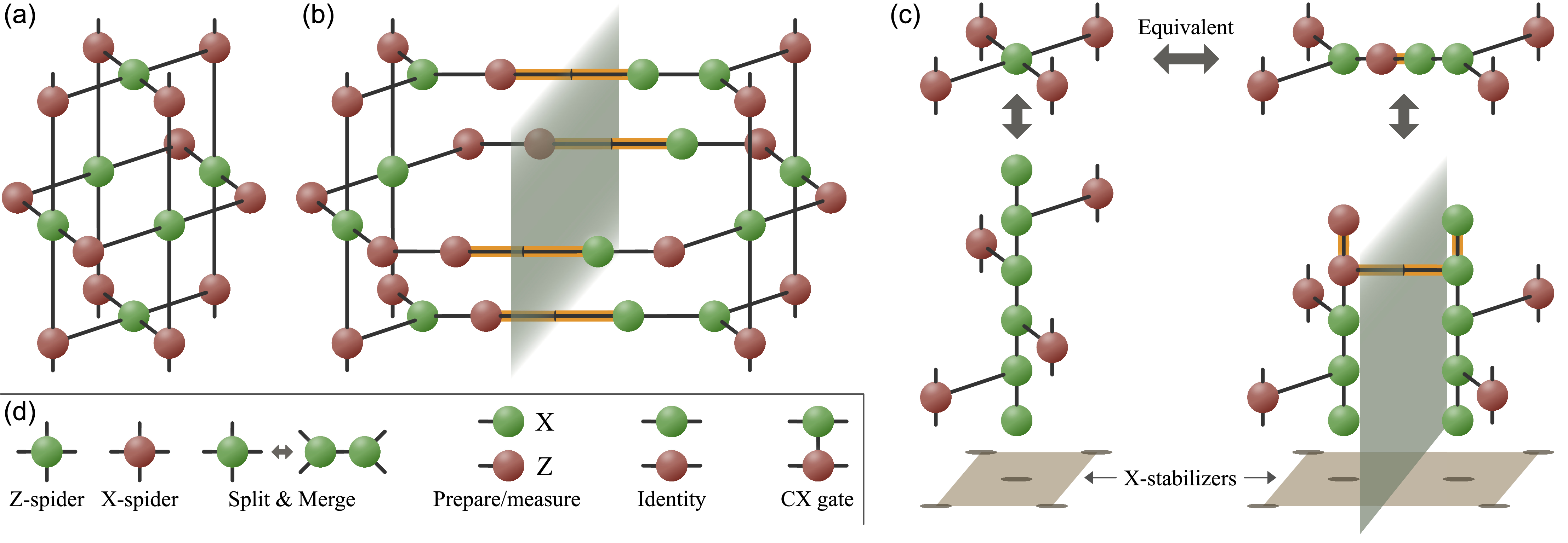}
    \caption{ZX diagrams for detection of phase flips in the surface code. (a) The simplest ZX diagram in which all spiders are connected to spiders of the opposite type. (b) At the interface between two modules, we split spiders in the diagram so that the number of legs crossing the interface is four. We then insert two-legged spiders of opposite type on each side of the interface to illustrate that links between the two sides of the interface can be implemented with CX gates as part of a Bell measurement. These links are highlighted in orange for clarity. (c) The ZX diagrams can be further expanded to give intuition about how they are implemented with quantum circuits on physical qubits. On the left is an expansion that produces the quantum circuit for measuring surface-code X-stabilizers in the bulk. On the right is the corresponding expansion for measuring the same stabilizer at the interface. (d) Elementary building blocks for ZX diagrams and their interpretations in terms of operations on physical qubits.}
    \label{fig:ZX}
\end{figure*}

Identifying detectors in the surface code is not trivial when teleported gates and Bell measurements are used. 
We therefore employ the language of ZX calculus~\cite{wetering_zx-calculus_2020}, which is useful for reasoning about error correction and fault tolerance~\cite{gidney_pair_2023, bombin_unifying_2024}.  
We use the convention that red vertices represent X-spiders and green vertices represent Z-spiders. 
Each spider is associated with a set of stabilizer generators. 
For an $n$-legged Z-spider, the generators are $\{X_1 X_2 \cdots X_n, Z_1 Z_i\}$ for all positive integers $i \leq n$. 
Analogously, the stabilizer generators associated with an $n$-legged X-spider are $\{Z_1 Z_2 \cdots Z_n, X_1 X_i\}$. 
Detectors are identified in a ZX diagram when the stabilizer generators associated with the spiders combine to form the identity. 
A ZX diagram that produces the surface-code detectors for phase flips is shown in \figpanel{fig:ZX}{a}. 
Each Z-spider (green) has a stabilizer $X_1X_2X_3X_4$, and each X-spider (red) has a stabilizer $X_1X_2$. 
Since each Z-spider is connected to an X-spider on all four legs and the network forms a closed cell, the stabilizers for the full structure combine to the identity. 
If a $Z$ error occurs on one of the qubits used to construct the ZX diagram in \figpanel{fig:ZX}{a}, this corresponds to inserting a two-legged Z-spider with a $\pi$ phase in the diagram. 
The addition of this spider changes the stabilizer product of the full structure from $I$ to $-I$. 
This change manifests as a flipped measurement outcome, which changes the parity of measurements in the detector. 

The transformation rules of ZX calculus allow spiders to be split and merged, which produces new but equivalent ZX diagrams. Equivalent diagrams have the same stabilizers and therefore represent the same detectors. 
When an interface needs to be introduced that separates two modules, we use this freedom to modify the ZX diagram as shown in \figpanel{fig:ZX}{b}. 
Spiders at the interface are split and new, phaseless two-legged spiders---which are equivalent to the identity---of opposite type are inserted on either side of the interface. 
The result is a diagram where four edges crossing the interface connect spider nodes of opposite type. 

The ZX diagrams in \figpanels{fig:ZX}{a}{b} can be interpreted as a series of operations on qubits through time. 
We associate time with the vertical direction in the diagrams.
In this picture, vertical connections between spider nodes represent qubit worldlines, and horizontal connections represent gates between qubits.
Specifically, each horizontal connection between an X-spider and a Z-spider represents a CX gate between qubits. 
According to our design criteria, a protocol should minimize the number of gates that cross the interface between modules. 
Detectors like the one shown in \figpanel{fig:ZX}{b} have four connections across the interface, but since each connection is part of two detectors, the average number of interface-crossing gates per detector is two. 
This is half as many as the benchmark protocol, and we conjecture that it is, in fact, optimal. 

In \figpanel{fig:ZX}{c}, we take the top part of the surface-code ZX diagrams and apply further spider splits to show how the diagram can be compiled into a quantum circuit. 
In the bulk, the diagram is compiled into the usual circuit for measuring stabilizers in the surface code: an ancilla qubit is prepared in the X basis, then undergoes CX gates with its neighboring data qubits before it is measured in the X basis. 
At the interface, the diagram is compiled into a circuit with two ancilla qubits that undergo CX gates with neighboring data qubits and an interface-crossing CX gate between the ancillae. 
The ZX diagrams do not specify the direction time flows, i.e., whether the interface-crossing CX gate should be applied before or after the local CX gates. 
In the diagram on the left in \figpanel{fig:ZX}{c}, it does not matter which way time flows since the diagram is invariant with respect to this choice.
At the interface, however, the choice gives two different interpretations.
If time is chosen to flow downward in \figpanel{fig:ZX}{c} so that the interface-crossing gate is applied before local gates, this produces the circuit in the measurement-teleportation protocol from Ref.~\cite{jacinto_network_2025}. 
If time is chosen to flow upward so that local gates are applied before the interface-crossing CX gate, this produces a circuit where the final operation on the ancilla qubits is a Bell measurement. 
Both circuits implement the surface code, but the Bell-measurement protocol is preferred because noise from the interface cannot propagate to the bulk.

Bell measurements produce twice as many measurement outcomes as single-ancilla measurements.
This increases the number of measurements in detectors at the interface. 
Detectors in the bulk of the surface code are formed from only two measurements: one measurement outcome from each of two subsequent rounds of syndrome measurement.
Detectors at the interface are formed from one measurement outcome from each of the four Bell measurements necessary to construct the ZX diagram in \figpanel{fig:ZX}{b}.  
An error at the interface will flip the outcome of the Bell measurement where the error occurred. 
Two detectors that share a measurement outcome which can be flipped by a single error are said to share a \textit{fault mechanism}. 
In the syndrome graph, detectors correspond to vertices and fault mechanisms correspond to edges. 
Bell measurements introduce two new edges for each vertex in the syndrome graph.
The valence of the syndrome graph at the interface is eight under phenomenological noise, compared to six for the bulk.
The new edges are in the same plane as the interface between modules; they are diagonal with both a space- and a time-like direction.
Edges from Bell measurements decrease the code distance at the interface if they are allowed to form uninterrupted chains in the syndrome graph.
This is similar to how hook errors under circuit-level noise form diagonal edges in the syndrome graph~\cite{jacinto_network_2025}. 
The orientation of diagonal edges due to hook errors is set by the order of gates in the syndrome-measurement circuit. 
This is true for diagonal edges due to Bell measurements as well.
Hence, the choice of syndrome-measurement circuit effectively determines the distance of the surface code at the interface between modules. 

\section{Alternating syndrome-measurement circuits}
\label{sec:circuit_wiggle}
\begin{figure}
    \centering
    \includegraphics[width=\linewidth]{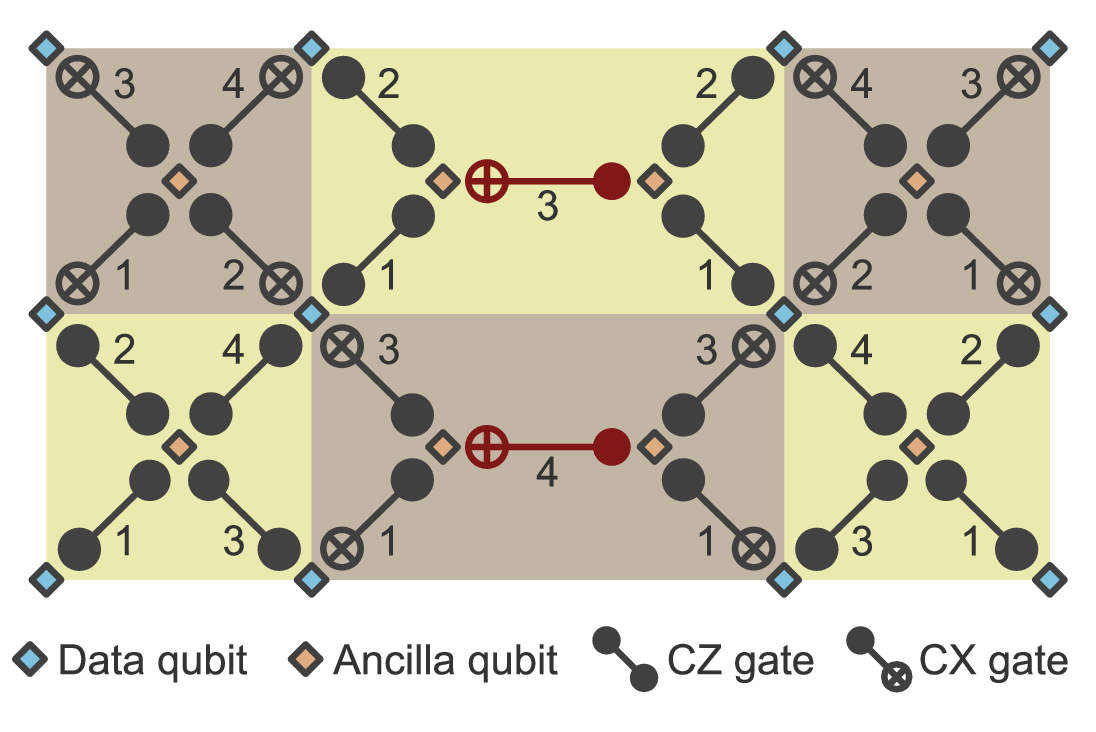}
    \caption{Syndrome-measurement circuits for a rotated surface code. Light (dark) plaquettes represent Z(X)-stabilizers. The syndrome-measurement circuit in the absence of an interface is shown on the square plaquettes. A zig-zag pattern is chosen to avoid $X$ errors propagating vertically and $Z$ errors propagating horizontally. The gate sequence in the Bell-measurement protocol is shown on rectangular plaquettes. Gates crossing the interface are colored red. The gate sequence must be alternated between rounds to avoid distance-reducing hook errors at the interface, as described in the main text.}
    \label{fig:gate-sequence}
\end{figure}

The syndrome-measurement circuit for rotated surface codes is designed to avoid propagation of errors through two-qubit gates that would reduce the distance of the code. 
The order of two-qubit gates between ancilla qubits and data qubits is required to follow a pattern in which the last two gates cannot propagate an error on the ancilla qubit to data qubits aligned parallel to a logical operator~\cite{tomita_low-distance_2014}. 
This restricts the viable choices of gate sequence to two, both of which follow a $\sf Z$-$\sf N$ pattern across a stabilizer.
We use CX gates for X-stabilizer measurements and CZ gates for Z-stabilizer measurements for clarity. 
One of the two distance-preserving measurement sequences is illustrated in \figref{fig:gate-sequence}. 
We assign the label A to this gate sequence. 
The alternative sequence is obtained by swapping the numbers $1\leftrightarrow 3$ and $2 \leftrightarrow 4$ for CX gates in the bulk, and $1\leftrightarrow 2$ and $3 \leftrightarrow 4$ for CZ gates in the bulk. 
We assign the label B to this gate sequence. 

The Bell-measurement protocol for lattice surgery does not require any changes to the standard syndrome-measurement circuit in the bulk. 
The gate sequence at the interface is chosen to match the circuit used to measure syndromes in the bulk, as shown in \figref{fig:gate-sequence}.
Interface-crossing CX gates that implement Bell measurements are applied after the local gates.
The gate sequence is symmetric about the interface to avoid qubits idling unnecessarily.
Note that the Bell measurement can be executed immediately after the local gates have been applied. 
The gate sequence at the interface that matches the alternative sequence in the bulk is obtained by swapping the numbers $1\leftrightarrow 3$ for local CX gates and $1\leftrightarrow 2$ for local CZ gates on plaquettes with an interface-crossing CX gate in \figref{fig:gate-sequence}. 

The full syndrome graph of a surface code under circuit-level noise is a three-dimensional graph that accounts for all Pauli errors. An example of a syndrome graph produced by the circuit in \figref{fig:gate-sequence} is provided in Appendix~\ref{sec:syndrome-graph}.
To aid understanding of the effect of Bell measurements on the syndrome graph, we will first assume that all local operations are perfect so that edges due to local errors in the syndrome graph can be removed. 
We are left with a syndrome graph where all edges represent Bell measurements at the interface. 
Since the interface has one spatial and one temporal dimension, the syndrome graph becomes two-dimensional. 

\begin{figure*}
    \centering
    \includegraphics[width=\linewidth]{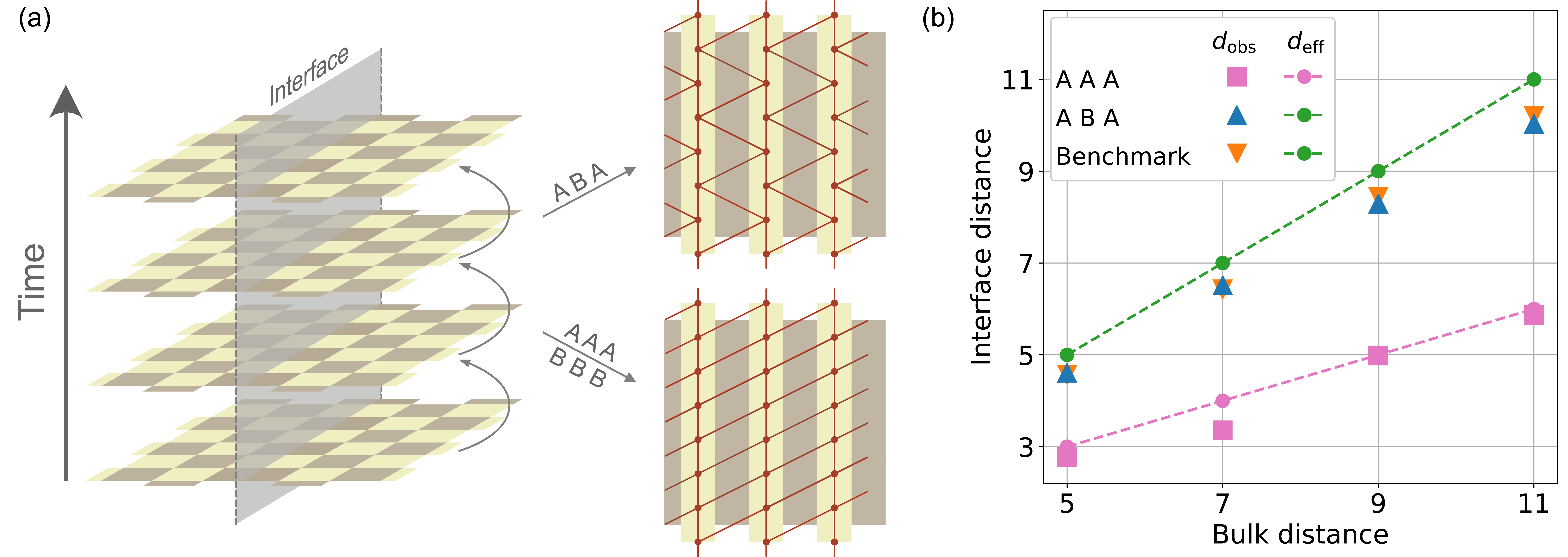}
    \caption{Effect of alternating the gate sequence between rounds of syndrome measurement. 
    Local gates and measurements are noiseless and gates crossing the interface are noisy. 
    (a) There are two possible choices for the gate sequence that do not cause hook errors in the bulk, which we label A and B. 
    Applying the same gate sequence for each round of syndrome measurement (AAA/BBB) produces a syndrome graph with parallel diagonal edges. 
    The parallel orientation of the edges represents physical errors aligning with the logical operator of the code to reduce the effective distance of the code at the interface. 
    Alternating the gate sequence between rounds of syndrome measurement (ABA) mitigates this effect by redirecting chains of physical errors. 
    The resulting syndrome graph has a zig-zag pattern of diagonal edges. 
    (b) Simulation of the effective code distance using repeating and alternating gate sequences for syndrome measurement. 
    Repeating the gate sequence every round of syndrome measurement effectively halves the distance of the code.
    Alternating gate sequences between rounds increases the effective code distance at the interface to $0.9d$, which matches the benchmark protocol. }
    \label{fig:syndrome-graph}
\end{figure*}

To allow for efficient decoding of surface codes, one needs to represent $Y$ errors as combinations of $X$ and $Z$ errors.
In that case, under standard noise model assumptions, the syndrome graph splits into two disconnected graphs, also known as the primal and dual syndrome graphs.
Typically, all edges in the primal (dual) syndrome graph represent bit-flip (phase-flip) errors.
In the remaining discussion, we consider only the primal syndrome graph.
All properties of the primal syndrome graph are shared by the dual. 
The primal syndrome graph at the interface between two modules for a distance-7 rotated surface code is shown in \figpanel{fig:syndrome-graph}{a}. 
The gate sequence in a round of syndrome measurements determines the orientation of the diagonal edges in the syndrome graph.
Repeating the same gate sequence every round leads to the syndrome graph shown at the bottom of \figpanel{fig:syndrome-graph}{a}.
Using alternating gate sequences between rounds produces the syndrome graph on the top.  
The syndrome graphs are overlaid on the surface-code detectors at the interface.
Paths on the syndrome graph that connect two disjoint X-type boundaries lead to undetectable logical bit flips.  
If a repeated gate schedule is used, such a path consists of $(d+1)/2$ edges.
Alternating the gate sequence changes the direction of half of the diagonal edges so that they form a zig-zag pattern. 
This alternation breaks the paths with $(d+1)/2$ edges.
The result is that the shortest path connecting two boundaries contains $d$ edges, which implies that the alternating gate sequence is distance-preserving. 
We stress that this is under the assumption that the bulk is noiseless. 

Alternating the gate sequence between rounds of syndrome measurement will not increase the effective code distance if Bell pairs are used to measure stabilizers at the interface, as in Refs.~\cite{shalby_optimized_2025, jacinto_network_2025}. 
Noise originating from Bell-state preparation inevitably propagates to all data qubits near the interface as gates are applied, regardless of the chosen gate sequence.

To confirm the different effective distances of the repeating and alternating gate sequences in the Bell-measurement protocol, we simulate the syndrome-measurement circuit with interface noise only using Stim~\cite{gidney_stim_2021}. 
For code sizes up to $d=11$, we use Stim to search for the shortest error chain that produces a logical error with trivial syndrome. 
The length of this error chain is the effective distance $d_\mathrm{eff}$ of the code. 
We find that $d_\mathrm{eff} = d$ for the alternating gate sequence, and $d_\mathrm{eff} = (d+1)/2$ for the repeating gate sequence. 

In addition to searching for the shortest undetectable logical error, we run Monte Carlo simulations of our circuits to determine the observed effective distance of the code, which we define by 
\begin{equation}
    p_\mathrm{L} \sim \left(\frac{p}{p_\mathrm{th}}\right)^{\frac{d_\mathrm{obs}+1}{2}},
    \label{eq:logical-error}
\end{equation}
where $p_\mathrm{L}$ is the logical error rate, $p_\mathrm{th}$ is the threshold error rate, and $p$ is the physical error rate. 
The observed effective distance can be extracted from the simulated $p_\mathrm{L} (p)$ for a fixed distance as 
\begin{equation}
    \frac{\partial \log p_\mathrm{L}}{\partial \log p} = \frac{d_\mathrm{obs} + 1}{2}.
    \label{eq:effective-distance}
\end{equation}
We approximate the left-hand side of \eqref{eq:effective-distance} as the difference in logical error rate when $p = \SI{0.1}{\percent}$ and $p = \SI{1}{\percent}$. 
To determine the logical error rate, we use a minimum-weight perfect-matching decoder~\cite{higgott_sparse_2025} to predict the logical state after $d$ rounds of syndrome measurement. 
Simulation results are shown in \figpanel{fig:syndrome-graph}{b}. 
Repeating the gate sequence between rounds halves the effective distance at the interface compared to the bulk, as expected. 
For the simulated code distances, the observed effective distance at the interface when gate sequences are alternated is $d_\mathrm{obs} = 0.9d$.
We also run a simulation using the benchmark protocol that produces no in-plane diagonal edges at the interface. 
We find no difference in effective distance at the interface between the alternating gate sequence in the Bell-measurement protocol and the benchmark protocol. 
The deviation of the observed effective distance from the length of the shortest error chain is likely due to finite-size effects, as \eqref{eq:logical-error} is only accurate in the low (physical) error rate regime~\cite{dennis_topological_2002}.

The effective distance at the interface due to errors in the Bell measurements alone will determine the logical error rate in the limit of vanishing local noise. 
However, significant local noise will be present in any implementation of surface codes on physical hardware. 
When a syndrome graph for the surface code is constructed with noise both in the bulk and at the interface, we find that the shortest error chain that causes an undetectable logical error is $3d/4$.
This reduction is a result of errors from local two-qubit gates that affect both a data qubit and a qubit used for Bell measurement.
To determine how strong this effect is, we resort to numerical studies in Section~\ref{sec:performance}.  

It is possible to restore the full distance at the interface by performing parity measurements on the ancilla qubits before the Bell measurements.
The parity measurements flag errors that contribute to the reduction in distance. However, interface-crossing parity measurements also introduce additional noisy gates and interfere with the gate schedule in the bulk. 
If gates across the interface have to be teleported by consuming a Bell pair, adding parity measurements to the syndrome-measurement circuit will increase the entanglement-generation rate required to support the protocol.

\section{Performance analysis}
\label{sec:performance}
\begin{figure}
    \centering
    \includegraphics[width=\linewidth]{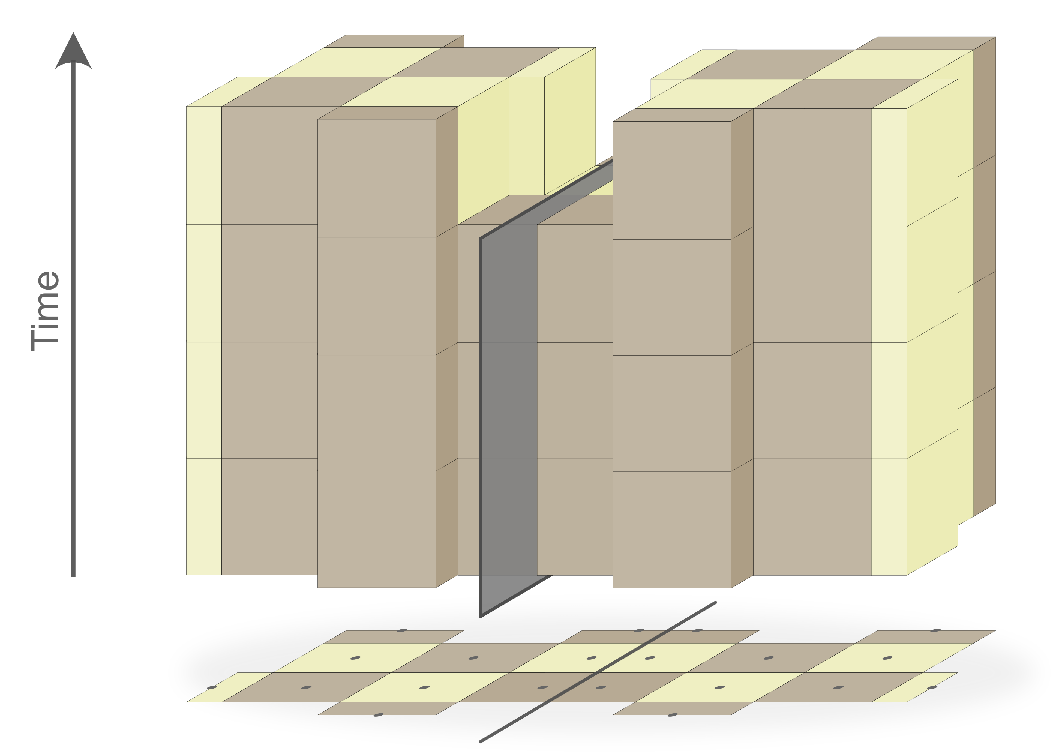}
    \caption{Preparation of a Bell state across an interface using a rotated surface code and lattice surgery. Dark boxes represent X-detectors and light boxes represent Z-detectors. At the bottom is a projection of the fault-tolerant channel onto a surface, where the physical layout of ancilla qubits used to extract syndrome information is shown. The interface between the two modules is represented by a dark sheet which bisects the channel.}
    \label{fig:fault-tolerant-channel}
\end{figure}

As an example application of the Bell-measurement protocol, we choose the task of establishing a logical Bell state shared by two separate modules.
If the modules are too far apart to execute CX or CZ gates between modules directly, these gates are teleported by consuming one ebit~\cite{bennett_concentrating_1996} in the form of a pre-shared physical Bell pair. 
For a code distance $d$, this consumes $d$ physical Bell pairs at the interface when the Bell-measurement protocol is used.  
The corresponding detector diagram in spacetime for code distance $d=3$ is shown in \figref{fig:fault-tolerant-channel}.
Two logical qubits are initially prepared in the state $\vert 0_\mathrm{L1}\rangle\vert 0_{\mathrm{L2}}\rangle$. 
A logical Bell state is obtained after measurement of the operator $X_{\mathrm{L1}} X_{\mathrm{L2}}$, which is performed by merging the two patches for $d$ rounds of syndrome extraction and then splitting the merged patch. 
The logical measurement outcome is revealed by combining the physical measurement outcomes of the new stabilizers in the merged region of the instrument~\cite{horsman_surface_2012, litinski_game_2019}. 
When Bell measurements are used to merge the patches at the interface, the measurement outcomes also determine the value of the operator $Z_{\mathrm{L1}} Z_{\mathrm{L2}}$, even in the absence of noise, as the states are initially in $\vert 0_\mathrm{L1}\rangle\vert 0_{\mathrm{L2}}\rangle$. 
This is similar to how the outcome of measurements in a qubit teleportation determine the Pauli correction that must be applied to the teleported qubit to recover the state.

We simulate the full circuit needed to construct the instrument under depolarizing noise using Stim~\cite{gidney_stim_2021}. 
Details about the error model are provided in Appendix~\ref{sec:error_model}.
We then sample errors from the circuit that prepares logical Bell states as shown in \figref{fig:fault-tolerant-channel} and decode the syndrome using a minimum-weight perfect-matching decoder~\cite{higgott_sparse_2025}. 
For comparison, we also simulate the circuit with the benchmark protocol that applies the standard surface-code gate sequence across the interface. 
The Bell-measurement protocol uses this sequence to measure stabilizers in the bulk (\figref{fig:gate-sequence}).
The local error rate $p_\mathrm{loc}$ is the error rate of operations that are executed within a module. 
These operations are physical qubit state preparation and measurement (SPAM), CX and CZ gates between nearest-neighbor physical qubits, and Hadamard gates. 
The link error rate $p_\mathrm{link}$ is the error rate of operations that are executed across the interface between modules. All such operations are either CX or CZ gates between two physical qubits in separate modules. 

\begin{figure*}
    \centering
    \includegraphics[width=\linewidth]{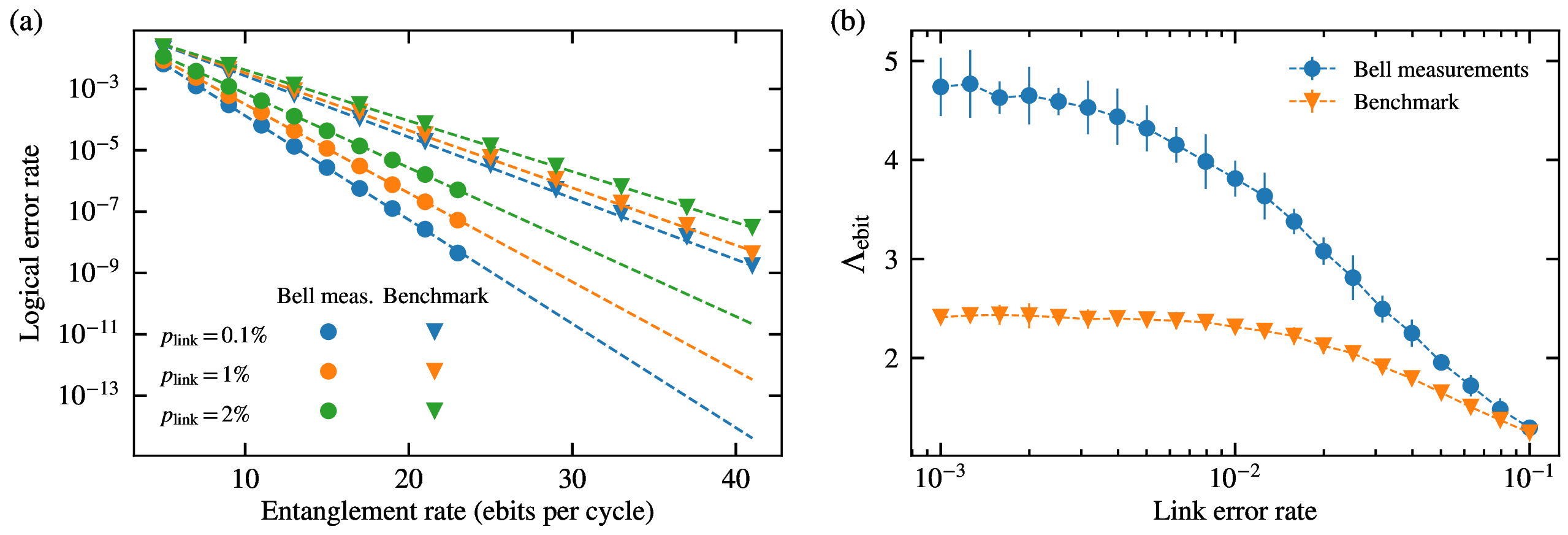}
    \caption{Sub-threshold performance comparison between the benchmark protocol and the Bell-measurement protocol we develop in this work. 
    (a) Logical error rate of the prepared Bell state as a function of the entanglement rate between modules. 
    Dashed lines represent exponential fits to the simulated data. 
    (b) Logical error suppression per ebit as a function of link error rate for the benchmark protocol and the Bell-measurement protocol. 
    Markers represent the average over distances up to 19; error bars show the standard deviation.}
    \label{fig:sim-results-teleportation}
\end{figure*}

We find that under uniform depolarizing noise $p_\mathrm{loc} = p_\mathrm{link}$, the error threshold is $p_\mathrm{th} = \SI{0.53}{\percent}$ for the benchmark protocol, while the Bell-measurement protocol has a threshold at $p_\mathrm{th} = \SI{0.52}{\percent}$. 
The logical error rate at threshold is three percentage points higher for the Bell-measurement protocol. 
Next, we set the local physical error rate to $p_{\mathrm{loc}} = \SI{0.1}{\percent}$ as we sweep the link error rate $p_{\mathrm{link}}$ from \SI{0.1}{\percent} to \SI{20}{\percent}. 
We find an error threshold at $p_\mathrm{link} = \SI{13.5}{\percent}$ for the benchmark protocol vs.~$p_\mathrm{link} = \SI{17.4}{\percent}$ for the Bell-measurement protocol. 
Details about the data from the threshold simulations are presented in Appendix~\ref{sec:thresholds}.
Using a teleported Bell measurement instead of a direct Bell measurement lowers the threshold value to $p_\mathrm{link} = \SI{16.9}{\percent}$. 
Unlike the Bell-measurement protocol, the benchmark protocol benefits from teleportation of gates when the interface noise is high. 
Errors on Bell pairs used for teleporting gates in Z-stabilizer measurements cannot propagate to data qubits, which reduces the noise experienced by the data qubits close to the interface. 
This is reflected in a link-error threshold for the teleported benchmark protocol at $p_\mathrm{link} = \SI{19.2}{\percent}$.

We now consider the case where the link noise is in the region $p_\mathrm{link} \in [\SI{0.1}{\percent}, \SI{10}{\percent}]$. 
This is the near-term feasible region where the link noise is below threshold for all protocols, yet still equal to or higher than the bulk noise. 
When high-fidelity inter-module entanglement is scarce, the Bell-measurement protocol is expected to achieve a stronger logical error suppression than the benchmark protocol because it requires only $d$ ebits, or Bell pairs, per round of syndrome measurement. 
This efficient use of ebits allows the Bell-measurement protocol to effectively double the distance at the interface relative to the benchmark protocol. 
In \figpanel{fig:sim-results-teleportation}{a}, we plot the achievable error rate of a logical Bell state as a function of the available entanglement rate between modules with the benchmark protocol and the Bell-measurement protocol, respectively, for different levels of link noise. 
We find that the Bell-measurement protocol benefits from its low consumption of ebits, with a saving of $\sim \SI{40}{\percent}$ relative to the benchmark protocol to achieve the same logical error rate. 
For example, by extrapolating our simulated error rates at $d \leq 23$ to higher distances, we find that to reach a logical error rate of $10^{-12}$ with a link error rate of $\SI{0.1}{\percent}$ ($\SI{1}{\percent}$) would require 35 (39) Bell pairs per round of syndrome measurement with the Bell-measurement protocol using a distance-35(39) surface code, compared to 57 (61) Bell pairs using a distance-29(31) surface code for the benchmark protocol. 

The suppression of the logical error rate with code distance is exponential. 
It is captured by the suppression factor $\Lambda_d = p_\mathrm{L}(d)/p_\mathrm{L}(d+2)$, which represents the drop in logical error rate when the code distance is increased by two.  
Similarly, we define the suppression factor $\Lambda_\mathrm{ebit}$ as the reduction in logical error rate when the number of ebits available per syndrome-measurement cycle increases by two. 
We plot the simulated value of $\Lambda_\mathrm{ebit}$ as a function of the link error rate $p_\mathrm{link}$ in \figpanel{fig:sim-results-teleportation}{b}. 
The Bell-measurement protocol offers stronger logical error suppression than the benchmark protocol for any level of interface noise in our simulations when access to ebits is limited. 
The suppression relative to the benchmark protocol is stronger for lower levels of interface noise and increases with the size of the code. 

\begin{figure*}
    \centering
    \includegraphics[width=\linewidth]{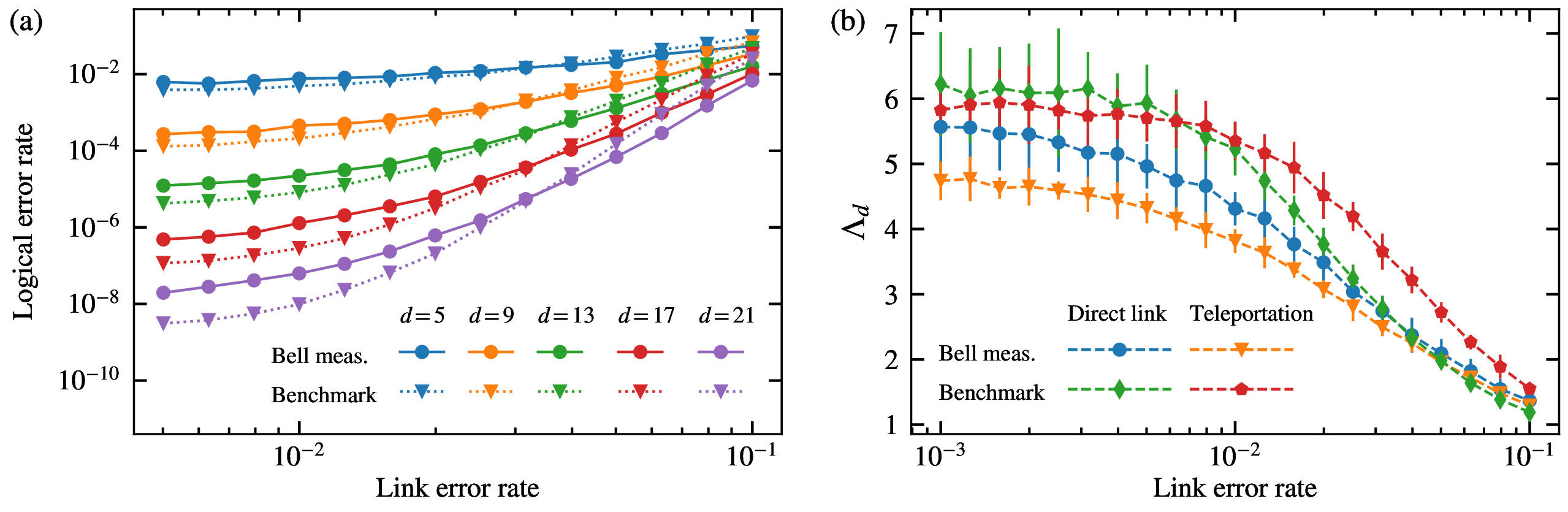}
    \caption{Sub-threshold performance when entanglement between modules is not a scarce resource. 
    (a) Logical error rate of the prepared Bell state as a function of link noise when gates are applied directly across the interface.
    (b) Logical error suppression per added distance as a function of link error rate for the syndrome-measurement circuits in this work.
    Markers represent the average over distances up to 19; error bars show the standard deviation.}
    \label{fig:sim-results}
\end{figure*}

We also consider the case where gates between modules are applied directly without teleportation. 
The simulated logical error rates as a function of code distance are shown in \figpanel{fig:sim-results}{a}. 
When $p_{\mathrm{link}} > \SI{3.4}{\percent}$, the Bell-measurement protocol outperforms the benchmark protocol despite the reduced distance at the interface. 
We believe this is due to the fact that the Bell-measurement protocol eliminates error propagation to qubits away from the interface, whereas the benchmark protocol applies noisy gates directly to data qubits close to the interface that take part in subsequent noise-propagating two-qubit gates. 
By decreasing the link noise below \SI{1}{\percent}, the logical error rate approaches a floor set by the error rate in the bulk for both protocols. 
This logical error floor is approximately a factor two higher for the Bell-measurement protocol. 
Adding parity checks to recover the full distance at the interface does not lower this floor, which suggests that the higher error floor is a consequence of the extra gates and measurements needed to perform Bell measurements, or the added edges in the syndrome graph at the interface that result from using Bell measurements, or a combination of these two effects.

We plot the simulated values of $\Lambda_d$ for the benchmark and Bell-measurement protocols in \figpanel{fig:sim-results}{b}. 
The benchmark protocol benefits significantly from teleportation of gates when $p_\mathrm{link} > \SI{1}{\percent}$. 
This agrees with the elevated threshold we found for the benchmark protocol and is due to noise confinement. 
The Bell-measurement protocol is maximally noise-confining by design.
The difference between direct and teleported gates is therefore negligible for this protocol at high levels of interface noise. 
As the link error rate decreases towards the bulk error rate, the relative importance of noise confinement decreases.
The overhead associated with teleportation means noise suppression for the teleportation protocols is not as strong as the direct link protocols for link noise levels close to that of the bulk. 
Suppressing logical errors by increasing the distance at low levels of interface noise is not as effective with the Bell-measurement protocol compared to the benchmark protocol. 
As a result, the Bell-measurement protocol typically requires an increase in distance at the interface by $\sim\SI{20}{\percent}$ relative to the benchmark protocol to reach the same logical error rate. 
This increased distance is the reason the Bell-measurement protocol saves only $\sim \SI{40}{\percent}$ on ebits compared to the benchmark protocol.

\section{Conclusion and outlook}
\label{sec:conclusion_outlook}
We have introduced a protocol for performing lattice surgery with Bell measurements at the interface between processor modules in a distributed quantum computer using the rotated surface code. 
The protocol is based on the insight that Bell measurements and Bell pairs are interchangeable in measurement-based quantum computing.
The Bell measurements provide two main advantages compared to Bell pairs used in other protocols. 
First, they strongly confine the elevated noise at the interface. 
Second, a protocol based on Bell measurements requires only $d$ ebits for gate teleportation for each round of syndrome measurement in a rotated surface code, compared to $2d-1$ for established protocols. 
This makes it particularly attractive for architectures that use gate teleportation between modules.
The performance of the protocol shows that it is competitive over a wide range of noise levels.
Due to its effective use of ebits, the Bell-measurement protocol offers a stronger logical error suppression per consumed ebit than alternative protocols, albeit at the cost of an increased number of physical qubits to achieve the same distance.
As an example, we find a $\SI{40}{\percent}$ reduction in entanglement rate needed to prepare a logical Bell state with an error rate of $10^{-12}$ between two modules with local physical error rate at $\SI{0.1}{\percent}$ and link error rates up to $\SI{1}{\percent}$.
However, there is still room for improvement because interface errors can combine with local errors to form undetectable logical errors of length $3d/4$ such that the physical qubit overhead to reach the same effective distance is increased in comparison to alternative schemes that require more shared entanglement, such as the protocols in Refs.~\cite{shalby_optimized_2025, jacinto_network_2025}.

We expect that the availability of high-fidelity Bell pairs will be a limiting factor for modular architectures for the foreseeable future, particularly for modules in separate cryostats or vacuum chambers.
Optically distributed entanglement of superconducting circuits in separate dilution refrigerators is yet to be demonstrated experimentally. 
Platforms such as trapped ions and neutral atoms have optical transitions that can be used to generate photon-mediated entanglement, but even here the rate of ebit generation has not exceeded \SI{200}{\hertz}~\cite{stephenson_high-rate_2020}.
We expect the Bell-measurement protocol to accelerate the development of modular, fault-tolerant quantum computers by effectively halving the entanglement-generation rate required to support inter-module lattice surgery. 
Our design principles can be applied to any quantum circuit that is distributed across processor modules, which can share only a limited amount of entanglement, thus cutting the cost of implementing such circuits in modular architectures. 

The effective distance of the rotated surface code we study in this paper depends on the gate sequence over two rounds of syndrome measurements. 
We have introduced a technique that alternates gate sequences between subsequent rounds to avoid distance-reducing errors at the interface between modules. 
This technique could have applications in the implementation of other quantum low-density parity-check codes that suffer from distance-reducing hook errors.

Future work could look to break the trade-off between low ebit consumption and full effective code distance that still exists in the protocol presented here. 
Another natural extension of this work is to study the impact of missing Bell pairs on code performance, as it is likely that hardware responsible for heralding Bell pairs will occasionally fail due to the probabilistic nature of heralding. 
Varying the noise model to identify regimes where different protocols provide a particular advantage over others could provide further insight. 

\begin{acknowledgments}

We thank Moritz Lange and Mats Granath for helpful discussions. 
T.H.\@ acknowledges the financial support from the Chalmers Excellence Initiative Nano and the Knut and Alice Wallenberg Foundation through the Wallenberg Centre for Quantum Technology (WACQT).
A.F.K.~acknowledges support from the Swedish Foundation for Strategic Research (grant numbers FFL21-0279 and FUS21-0063), the Horizon Europe programme HORIZON-CL4-2022-QUANTUM-01-SGA via the project 101113946 OpenSuperQPlus100, and from the Knut and Alice Wallenberg Foundation through the Wallenberg Centre for Quantum Technology (WACQT). R.V.L. acknowledges support from the Knut and Alice Wallenberg foundation through the Wallenberg Centre for Quantum Technology (WACQT), from the European Research Council via Starting Grant 948265, and from the Swedish Foundation for Strategic Research via grant FFL21-0039.
Numerical simulations were performed using resources at the Chalmers Centre for Computational Science and Engineering (C3SE) under project 2025/1-6.

\end{acknowledgments}

\input{supplemental}

\clearpage

\bibliography{lattice_surgery}

\end{document}

%% file: supplemental.tex
\appendix

\section{Bell measurements}
\label{sec:Bell_circuits}

\begin{figure}
    \centering
    \includegraphics[width=\linewidth]{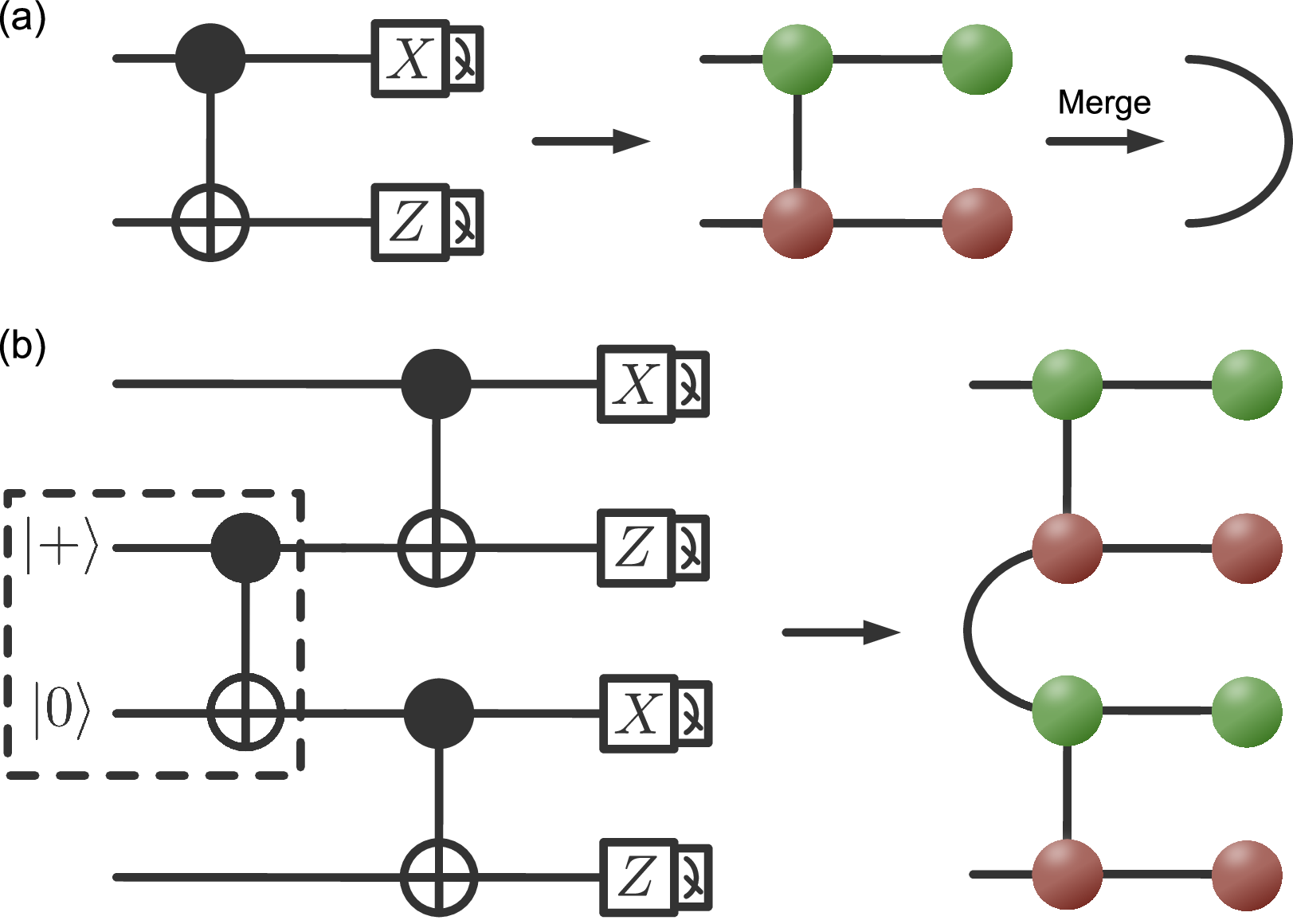}
    \caption{Bell-measurement circuits and their representation as ZX diagrams. (a) A standard Bell measurement on two qubits can be implemented with one CX gate followed by single-qubit measurements. 
    The control qubit is measured in the X basis to determine the observable $X_1X_2$, and the target qubit is measured in the Z basis to determine the observable $Z_1Z_2$.
    (b) A teleported Bell measurement requires an initial preparation of a Bell pair (dashed box). }
    \label{fig:circuits}
\end{figure}

Bell measurements are joint measurements on two qubits in the Bell basis, which are the four Bell states that are simultaneous eigenstates of the Pauli operators $X_1X_2$ and $Z_1Z_2$. 
The Bell basis is a complete basis for the Hilbert space of two qubits.
The quantum circuit and corresponding ZX diagram for a Bell measurement are shown in \figref{fig:circuits}.
The circuit for a Bell measurement is simply the inverse of the circuit that prepares a Bell state between two qubits. 
Bell measurements can be implemented deterministically in platforms with access to entangling gates, e.g. superconducting qubits, atoms and ions, because the circuit in \figref{fig:circuits} distinguishes all four Bell states. 

In the ZX diagram language, Bell measurements and Bell states can be represented by a single piece of wire, as illustrated in \figpanel{fig:circuits}{a} by converting the circuit for a Bell measurement into the corresponding ZX diagram using the building blocks in \figpanel{fig:ZX}{d}.
This representation is particularly helpful when we are looking for possible implementations of fault-tolerant protocols described by ZX diagrams. 
Interfaces between modules can be modeled as planes cutting through the diagrams, and wires that cross planes can be interpreted as Bell measurements on qubits between modules. 
The ZX calculus allows spiders to be split and merged to find planes that intersect as few wires as possible. 
These planes represent optimal divisions of an operation described by the ZX diagram across modules with the respect to the amount of entanglement that is needed to implement it.
We implement this strategy on the ZX diagram that describes the surface code in this work, but the approach is applicable to any quantum circuit that must be divided among processor modules that can share a limited amount of entanglement. 

\section{Error model}
\label{sec:error_model}

\begin{table*}[b]
    \centering
    \renewcommand{\arraystretch}{1.2} 
    \setlength{\tabcolsep}{10pt}
    \begin{tabular}{lccccc}
        \toprule
        \textbf{Operation} & \textbf{Idling} & \textbf{Hadamard} & \textbf{CX/CZ} & \textbf{SPAM} & \textbf{Interface CX/CZ} \\
        Stim instruction & \texttt{DEPOLARIZE1} & \texttt{DEPOLARIZE1} & \texttt{DEPOLARIZE2} & \texttt{X\_ERROR} & \texttt{DEPOLARIZE2} \\
        Uniform noise & $p_\mathrm{loc}$ & $p_\mathrm{loc}$ & $p_\mathrm{loc}$ & $p_\mathrm{loc}$ & $p_\mathrm{link} = p_\mathrm{loc}$ \\
        Elevated link noise & \SI{0.1}{\percent} & \SI{0.1}{\percent} & \SI{0.1}{\percent} & \SI{0.1}{\percent} & $p_\mathrm{link}$ \\
    \end{tabular}
    \caption{Summary of the error model used in the simulations.}
    \label{tab:error_model}
\end{table*}

We use a primarily depolarizing error model for the circuit-level simulations.
Each single-qubit gate (including the identity gate) is followed by a depolarizing channel which, with probability $p$, applies one Pauli operator $P \in \{ X, Y, Z\}$ chosen at random. 
That is, the operator $X$ is applied with probability $p/3$, the operator $Z$ is applied with probability $p/3$, or the operator $Y$ is applied with probability $p/3$. 
The probability of no error is then $1-p$. Each two-qubit gate is followed by a two-qubit depolarizing channel which, with probability $p$, applies $P_A \otimes P_B \neq I\otimes I$ where $P_{A, B} \in \{X, Y, Z, I\}$. 
That is, each of the Pauli product operators have a probability $p/15$ of being applied, and the probability of no error is $1-p$. 
All measurements are done in the Z basis; measurements in the X basis are implemented with a Hadamard followed by a Z-basis measurement. 
Each measurement outcome can return the wrong result with probability set by the state-preparation and measurement (SPAM) error rate.  
All qubits are reset to $\vert 0\rangle$ after measurement and a bit-flip error may occur after the reset. 
The error model and noise levels are summarized in \tabref{tab:error_model}. 

When a gate needs to be teleported, we first prepare a Bell pair between the two modules using perfect operations.
The resulting state is
\begin{equation}
    \vert\Phi^+\rangle = \frac{1}{\sqrt{2}}\left(\vert 00\rangle + \vert 11\rangle\right).
\end{equation}
Ideal preparation is followed by a two-qubit depolarizing channel with noise strength $p_\mathrm{link}$. 
After the channel, the state shared between the modules is
\begin{equation}
    \rho = \sum\limits_{i=0}^{15} K_i  \vert\Phi^+\rangle \langle\Phi^+\vert K_i^\dagger,
\end{equation}
where $\{K_i\} = \{\sqrt{1-p_\mathrm{link}}\ I_A \otimes I_B , \sqrt{p_\mathrm{link}/15}\ P_A \otimes P_B\}$ are the Kraus operators for the depolarizing channel. 
The fidelity of the resulting Bell pair is
\begin{equation}
    F = \langle \Phi^+ \vert \rho \vert \Phi^+\rangle = 1-p_\mathrm{link} + \frac{3p_\mathrm{link}}{15} = 1-\frac{4p_\mathrm{link}}{5}.
\end{equation}

\section{Syndrome graph}
\label{sec:syndrome-graph}
A syndrome graph is a graphical representation of the error syndromes obtained from the detectors of an error-correcting code. 
Detectors are represented by vertices in the graph, and faults that can cause a detector to flip correspond to edges incident on a vertex. 
Under circuit-level noise, the syndrome graph for the surface code is a (2+1)-dimensional spacetime graph where edges represent gate errors, idling errors, or errors on qubits during state preparation or measurement.

\begin{figure}
    \centering
    \includegraphics[width=\linewidth]{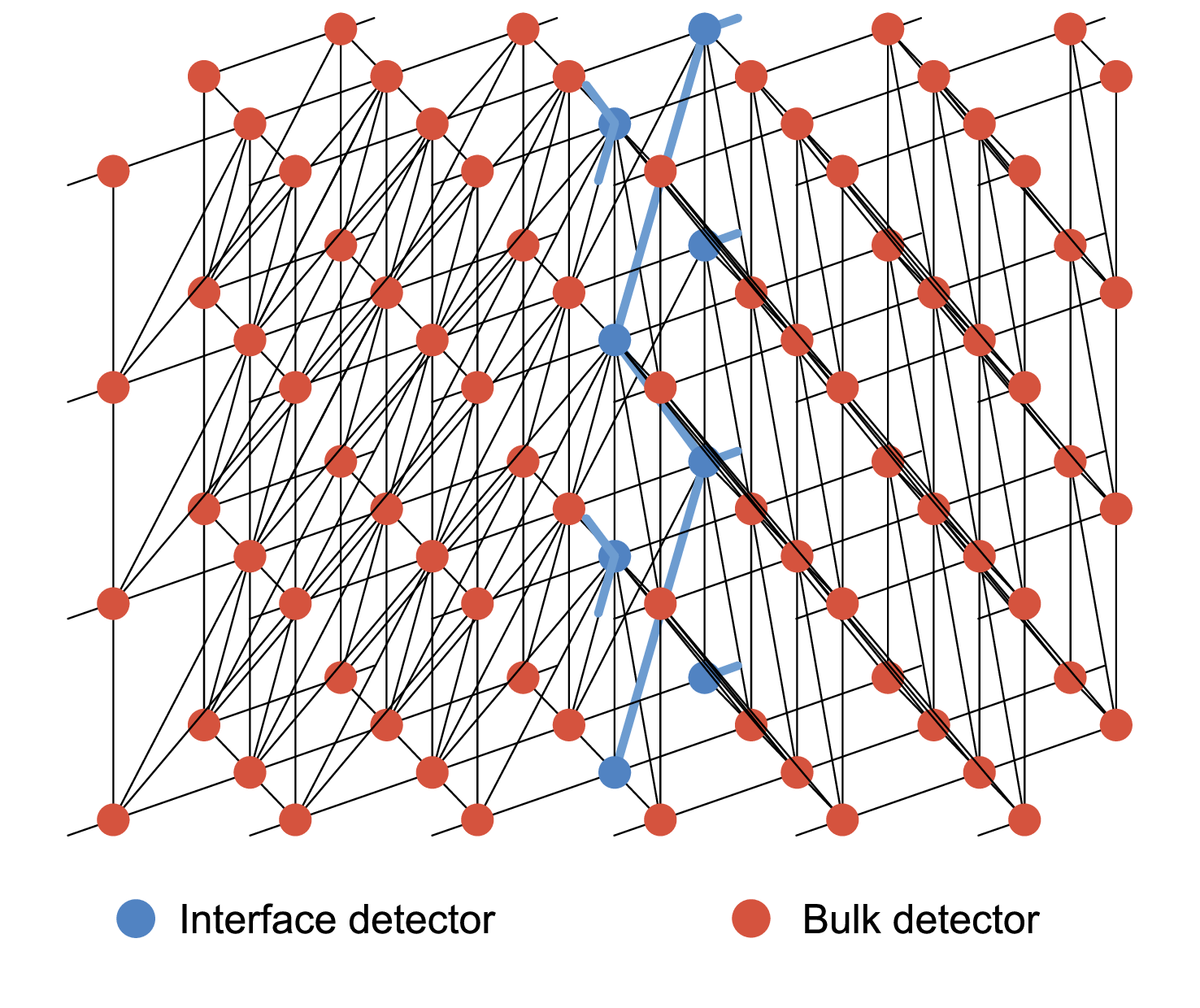}
    \caption{Full syndrome graph of two merged distance-5 rotated surface codes under circuit-level noise with an interface between modules in the center. Note that time-like boundaries are not included. Edges that correspond to faults at the interface are highlighted.}
    \label{fig:full_syndrome_graph}
\end{figure}

The structure of the syndrome graph depends on the gate sequence used to measure detectors. 
The circuit that we use to extract error syndromes in this work produces a syndrome graph of the form shown in \figref{fig:full_syndrome_graph}.
The syndrome graph is generally an undirected simple graph where each vertex has degree twelve.
However, the introduction of an alternative gate sequence at the interface produces vertices close to the interface that have degree eleven.
The gate sequence is symmetric about the interface, and hence the syndrome graph becomes symmetric about the interface as well. 

\section{Threshold simulations}
\label{sec:thresholds}

\begin{figure*}
    \centering
    \includegraphics[width=\linewidth]{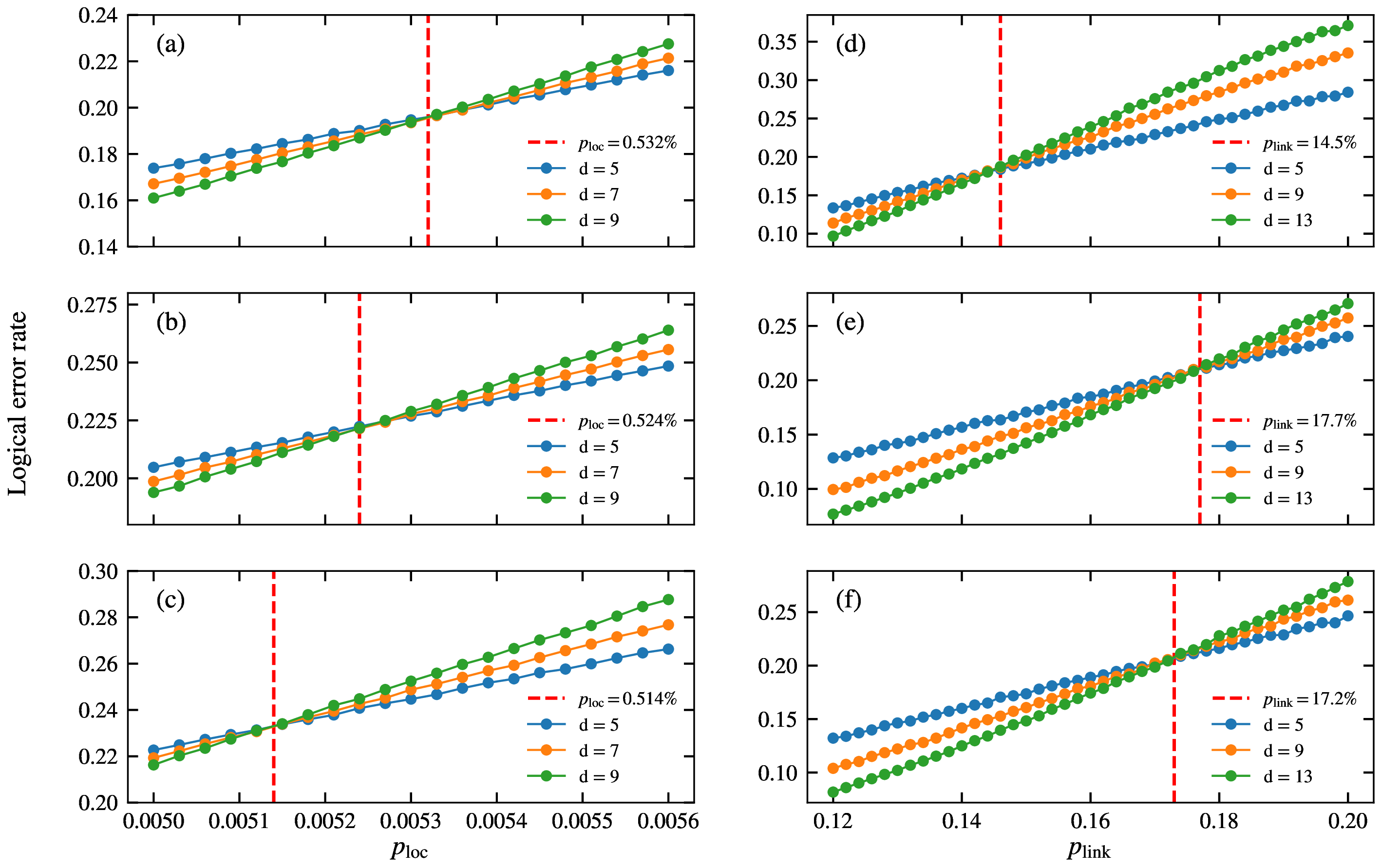}
    \caption{Data from threshold simulations. (a) The benchmark protocol with direct link under uniform noise. 
    (b) The Bell-measurement protocol with direct link under uniform noise.
    (c) The Bell-measurement protocol with teleported gates under uniform noise.
    (d) The benchmark protocol with direct link under elevated link noise.
    (e) The Bell-measurement protocol with direct link under elevated link noise. 
    (f) The Bell-measurement protocol with teleported gates under elevated link noise.}
    \label{fig:thresholds}
\end{figure*}

The results from simulations identifying the thresholds for the benchmark and Bell-measurements protocols under uniform noise are plotted in \figpanels{fig:thresholds}{a}{c}. 
We find that the logical error rate close to the threshold is slightly higher for the Bell-measurement protocol than the benchmark protocol, and the threshold value is slightly lower. 
This is expected due to the added complexity and local reduction in effective distance due to the Bell measurements, which are advantageous only when there is a noisy interface between modules.  

We also consider a situation where the local error rate in both modules is set to $p_\mathrm{loc} = \SI{0.1}{\percent}$ but the link noise is elevated. 
This is a much more realistic scenario with near-term hardware constraints. 
The simulation results are shown in \figpanels{fig:thresholds}{d}{f}. 
As explained in the main text, the Bell-measurement protocol is more resilient to high link noise than the benchmark protocol when gates are applied across the interface without teleportation.
The fact that the Bell measurements fully confine the link noise to the ancilla qubits at the interface outweighs the reduction in effective distance. 
This is no longer the case when gates are teleported. 
Gate teleportation protects the data qubits from some of the link noise.
Because the Bell-measurement protocol is intrinsically fully noise-confining, performance is slightly compromised by the overhead required for teleportation.

%% file: main.bbl
\begin{thebibliography}{51}%
\makeatletter
\providecommand \@ifxundefined [1]{%
 \@ifx{#1\undefined}
}%
\providecommand \@ifnum [1]{%
 \ifnum #1\expandafter \@firstoftwo
 \else \expandafter \@secondoftwo
 \fi
}%
\providecommand \@ifx [1]{%
 \ifx #1\expandafter \@firstoftwo
 \else \expandafter \@secondoftwo
 \fi
}%
\providecommand \natexlab [1]{#1}%
\providecommand \enquote  [1]{``#1''}%
\providecommand \bibnamefont  [1]{#1}%
\providecommand \bibfnamefont [1]{#1}%
\providecommand \citenamefont [1]{#1}%
\providecommand \href@noop [0]{\@secondoftwo}%
\providecommand \href [0]{\begingroup \@sanitize@url \@href}%
\providecommand \@href[1]{\@@startlink{#1}\@@href}%
\providecommand \@@href[1]{\endgroup#1\@@endlink}%
\providecommand \@sanitize@url [0]{\catcode `\\12\catcode `\$12\catcode `\&12\catcode `\#12\catcode `\^12\catcode `\_12\catcode `\%12\relax}%
\providecommand \@@startlink[1]{}%
\providecommand \@@endlink[0]{}%
\providecommand \url  [0]{\begingroup\@sanitize@url \@url }%
\providecommand \@url [1]{\endgroup\@href {#1}{\urlprefix }}%
\providecommand \urlprefix  [0]{URL }%
\providecommand \Eprint [0]{\href }%
\providecommand \doibase [0]{https://doi.org/}%
\providecommand \selectlanguage [0]{\@gobble}%
\providecommand \bibinfo  [0]{\@secondoftwo}%
\providecommand \bibfield  [0]{\@secondoftwo}%
\providecommand \translation [1]{[#1]}%
\providecommand \BibitemOpen [0]{}%
\providecommand \bibitemStop [0]{}%
\providecommand \bibitemNoStop [0]{.\EOS\space}%
\providecommand \EOS [0]{\spacefactor3000\relax}%
\providecommand \BibitemShut  [1]{\csname bibitem#1\endcsname}%
\let\auto@bib@innerbib\@empty
\bibitem [{\citenamefont {Shor}(1997)}]{shor_polynomial-time_1997}%
  \BibitemOpen
  \bibfield  {author} {\bibinfo {author} {\bibfnamefont {P.~W.}\ \bibnamefont {Shor}},\ }\href {https://doi.org/10.1137/S0097539795293172} {\bibfield  {journal} {\bibinfo  {journal} {SIAM Journal on Computing}\ }\textbf {\bibinfo {volume} {26}},\ \bibinfo {pages} {1484} (\bibinfo {year} {1997})}\BibitemShut {NoStop}%
\bibitem [{\citenamefont {Gisin}\ and\ \citenamefont {Thew}(2007)}]{Gisin2007}%
  \BibitemOpen
  \bibfield  {author} {\bibinfo {author} {\bibfnamefont {N.}~\bibnamefont {Gisin}}\ and\ \bibinfo {author} {\bibfnamefont {R.}~\bibnamefont {Thew}},\ }\href {https://doi.org/10.1038/nphoton.2007.22} {\bibfield  {journal} {\bibinfo  {journal} {Nature Photonics}\ }\textbf {\bibinfo {volume} {1}},\ \bibinfo {pages} {165} (\bibinfo {year} {2007})}\BibitemShut {NoStop}%
\bibitem [{\citenamefont {Dalzell}\ \emph {et~al.}(2023)\citenamefont {Dalzell}, \citenamefont {McArdle}, \citenamefont {Berta}, \citenamefont {Bienias}, \citenamefont {Chen}, \citenamefont {Gily{\'{e}}n}, \citenamefont {Hann}, \citenamefont {Kastoryano}, \citenamefont {Khabiboulline}, \citenamefont {Kubica}, \citenamefont {Salton}, \citenamefont {Wang},\ and\ \citenamefont {Brand{\~{a}}o}}]{Dalzell2023}%
  \BibitemOpen
  \bibfield  {author} {\bibinfo {author} {\bibfnamefont {A.~M.}\ \bibnamefont {Dalzell}}, \bibinfo {author} {\bibfnamefont {S.}~\bibnamefont {McArdle}}, \bibinfo {author} {\bibfnamefont {M.}~\bibnamefont {Berta}}, \bibinfo {author} {\bibfnamefont {P.}~\bibnamefont {Bienias}}, \bibinfo {author} {\bibfnamefont {C.-F.}\ \bibnamefont {Chen}}, \bibinfo {author} {\bibfnamefont {A.}~\bibnamefont {Gily{\'{e}}n}}, \bibinfo {author} {\bibfnamefont {C.~T.}\ \bibnamefont {Hann}}, \bibinfo {author} {\bibfnamefont {M.~J.}\ \bibnamefont {Kastoryano}}, \bibinfo {author} {\bibfnamefont {E.~T.}\ \bibnamefont {Khabiboulline}}, \bibinfo {author} {\bibfnamefont {A.}~\bibnamefont {Kubica}}, \bibinfo {author} {\bibfnamefont {G.}~\bibnamefont {Salton}}, \bibinfo {author} {\bibfnamefont {S.}~\bibnamefont {Wang}},\ and\ \bibinfo {author} {\bibfnamefont {F.~G. S.~L.}\ \bibnamefont {Brand{\~{a}}o}},\ }\href@noop {} {\bibinfo {title} {{Quantum algorithms: A survey of applications and end-to-end complexities}}} (\bibinfo {year} {2023}),\
  \Eprint {https://arxiv.org/abs/2310.03011} {arXiv:2310.03011} \BibitemShut {NoStop}%
\bibitem [{\citenamefont {Huang}\ \emph {et~al.}(2025)\citenamefont {Huang}, \citenamefont {Choi}, \citenamefont {McClean},\ and\ \citenamefont {Preskill}}]{huang_vast_2025}%
  \BibitemOpen
  \bibfield  {author} {\bibinfo {author} {\bibfnamefont {H.-Y.}\ \bibnamefont {Huang}}, \bibinfo {author} {\bibfnamefont {S.}~\bibnamefont {Choi}}, \bibinfo {author} {\bibfnamefont {J.~R.}\ \bibnamefont {McClean}},\ and\ \bibinfo {author} {\bibfnamefont {J.}~\bibnamefont {Preskill}},\ }\href@noop {} {\bibinfo {title} {The vast world of quantum advantage}} (\bibinfo {year} {2025}),\ \Eprint {https://arxiv.org/abs/2508.05720} {arXiv:2508.05720} \BibitemShut {NoStop}%
\bibitem [{\citenamefont {Gu}\ \emph {et~al.}(2017)\citenamefont {Gu}, \citenamefont {Kockum}, \citenamefont {Miranowicz}, \citenamefont {Liu},\ and\ \citenamefont {Nori}}]{Gu2017}%
  \BibitemOpen
  \bibfield  {author} {\bibinfo {author} {\bibfnamefont {X.}~\bibnamefont {Gu}}, \bibinfo {author} {\bibfnamefont {A.~F.}\ \bibnamefont {Kockum}}, \bibinfo {author} {\bibfnamefont {A.}~\bibnamefont {Miranowicz}}, \bibinfo {author} {\bibfnamefont {Y.-X.}\ \bibnamefont {Liu}},\ and\ \bibinfo {author} {\bibfnamefont {F.}~\bibnamefont {Nori}},\ }\href {https://doi.org/10.1016/j.physrep.2017.10.002} {\bibfield  {journal} {\bibinfo  {journal} {Physics Reports}\ }\textbf {\bibinfo {volume} {718--719}},\ \bibinfo {pages} {1} (\bibinfo {year} {2017})}\BibitemShut {NoStop}%
\bibitem [{\citenamefont {Blais}\ \emph {et~al.}(2021)\citenamefont {Blais}, \citenamefont {Grimsmo}, \citenamefont {Girvin},\ and\ \citenamefont {Wallraff}}]{Blais2021}%
  \BibitemOpen
  \bibfield  {author} {\bibinfo {author} {\bibfnamefont {A.}~\bibnamefont {Blais}}, \bibinfo {author} {\bibfnamefont {A.~L.}\ \bibnamefont {Grimsmo}}, \bibinfo {author} {\bibfnamefont {S.~M.}\ \bibnamefont {Girvin}},\ and\ \bibinfo {author} {\bibfnamefont {A.}~\bibnamefont {Wallraff}},\ }\href {https://doi.org/10.1103/RevModPhys.93.025005} {\bibfield  {journal} {\bibinfo  {journal} {Reviews of Modern Physics}\ }\textbf {\bibinfo {volume} {93}},\ \bibinfo {pages} {025005} (\bibinfo {year} {2021})}\BibitemShut {NoStop}%
\bibitem [{\citenamefont {Krinner}\ \emph {et~al.}(2019)\citenamefont {Krinner}, \citenamefont {Storz}, \citenamefont {Kurpiers}, \citenamefont {Magnard}, \citenamefont {Heinsoo}, \citenamefont {Keller}, \citenamefont {Lütolf}, \citenamefont {Eichler},\ and\ \citenamefont {Wallraff}}]{krinner_engineering_2019}%
  \BibitemOpen
  \bibfield  {author} {\bibinfo {author} {\bibfnamefont {S.}~\bibnamefont {Krinner}}, \bibinfo {author} {\bibfnamefont {S.}~\bibnamefont {Storz}}, \bibinfo {author} {\bibfnamefont {P.}~\bibnamefont {Kurpiers}}, \bibinfo {author} {\bibfnamefont {P.}~\bibnamefont {Magnard}}, \bibinfo {author} {\bibfnamefont {J.}~\bibnamefont {Heinsoo}}, \bibinfo {author} {\bibfnamefont {R.}~\bibnamefont {Keller}}, \bibinfo {author} {\bibfnamefont {J.}~\bibnamefont {Lütolf}}, \bibinfo {author} {\bibfnamefont {C.}~\bibnamefont {Eichler}},\ and\ \bibinfo {author} {\bibfnamefont {A.}~\bibnamefont {Wallraff}},\ }\href {https://doi.org/10.1140/epjqt/s40507-019-0072-0} {\bibfield  {journal} {\bibinfo  {journal} {EPJ Quantum Technology}\ }\textbf {\bibinfo {volume} {6}},\ \bibinfo {pages} {1} (\bibinfo {year} {2019})}\BibitemShut {NoStop}%
\bibitem [{\citenamefont {Mohseni}\ \emph {et~al.}(2025)\citenamefont {Mohseni}, \citenamefont {Scherer}, \citenamefont {Johnson}, \citenamefont {Wertheim}, \citenamefont {Otten}, \citenamefont {Aadit}, \citenamefont {Alexeev}, \citenamefont {Bresniker}, \citenamefont {Camsari}, \citenamefont {Chapman}, \citenamefont {Chatterjee}, \citenamefont {Dagnew}, \citenamefont {Esposito}, \citenamefont {Fahim}, \citenamefont {Fiorentino}, \citenamefont {Gajjar}, \citenamefont {Khalid}, \citenamefont {Kong}, \citenamefont {Kulchytskyy}, \citenamefont {Kyoseva}, \citenamefont {Li}, \citenamefont {Lott}, \citenamefont {Markov}, \citenamefont {McDermott}, \citenamefont {Pedretti}, \citenamefont {Rao}, \citenamefont {Rieffel}, \citenamefont {Silva}, \citenamefont {Sorebo}, \citenamefont {Spentzouris}, \citenamefont {Steiner}, \citenamefont {Torosov}, \citenamefont {Venturelli}, \citenamefont {Visser}, \citenamefont {Webb}, \citenamefont {Zhan}, \citenamefont {Cohen}, \citenamefont {Ronagh}, \citenamefont {Ho},
  \citenamefont {Beausoleil},\ and\ \citenamefont {Martinis}}]{mohseni_how_2025}%
  \BibitemOpen
  \bibfield  {author} {\bibinfo {author} {\bibfnamefont {M.}~\bibnamefont {Mohseni}}, \bibinfo {author} {\bibfnamefont {A.}~\bibnamefont {Scherer}}, \bibinfo {author} {\bibfnamefont {K.~G.}\ \bibnamefont {Johnson}}, \bibinfo {author} {\bibfnamefont {O.}~\bibnamefont {Wertheim}}, \bibinfo {author} {\bibfnamefont {M.}~\bibnamefont {Otten}}, \bibinfo {author} {\bibfnamefont {N.~A.}\ \bibnamefont {Aadit}}, \bibinfo {author} {\bibfnamefont {Y.}~\bibnamefont {Alexeev}}, \bibinfo {author} {\bibfnamefont {K.~M.}\ \bibnamefont {Bresniker}}, \bibinfo {author} {\bibfnamefont {K.~Y.}\ \bibnamefont {Camsari}}, \bibinfo {author} {\bibfnamefont {B.}~\bibnamefont {Chapman}}, \bibinfo {author} {\bibfnamefont {S.}~\bibnamefont {Chatterjee}}, \bibinfo {author} {\bibfnamefont {G.~A.}\ \bibnamefont {Dagnew}}, \bibinfo {author} {\bibfnamefont {A.}~\bibnamefont {Esposito}}, \bibinfo {author} {\bibfnamefont {F.}~\bibnamefont {Fahim}}, \bibinfo {author} {\bibfnamefont {M.}~\bibnamefont {Fiorentino}}, \bibinfo {author} {\bibfnamefont
  {A.}~\bibnamefont {Gajjar}}, \bibinfo {author} {\bibfnamefont {A.}~\bibnamefont {Khalid}}, \bibinfo {author} {\bibfnamefont {X.}~\bibnamefont {Kong}}, \bibinfo {author} {\bibfnamefont {B.}~\bibnamefont {Kulchytskyy}}, \bibinfo {author} {\bibfnamefont {E.}~\bibnamefont {Kyoseva}}, \bibinfo {author} {\bibfnamefont {R.}~\bibnamefont {Li}}, \bibinfo {author} {\bibfnamefont {P.~A.}\ \bibnamefont {Lott}}, \bibinfo {author} {\bibfnamefont {I.~L.}\ \bibnamefont {Markov}}, \bibinfo {author} {\bibfnamefont {R.~F.}\ \bibnamefont {McDermott}}, \bibinfo {author} {\bibfnamefont {G.}~\bibnamefont {Pedretti}}, \bibinfo {author} {\bibfnamefont {P.}~\bibnamefont {Rao}}, \bibinfo {author} {\bibfnamefont {E.}~\bibnamefont {Rieffel}}, \bibinfo {author} {\bibfnamefont {A.}~\bibnamefont {Silva}}, \bibinfo {author} {\bibfnamefont {J.}~\bibnamefont {Sorebo}}, \bibinfo {author} {\bibfnamefont {P.}~\bibnamefont {Spentzouris}}, \bibinfo {author} {\bibfnamefont {Z.}~\bibnamefont {Steiner}}, \bibinfo {author} {\bibfnamefont
  {B.}~\bibnamefont {Torosov}}, \bibinfo {author} {\bibfnamefont {D.}~\bibnamefont {Venturelli}}, \bibinfo {author} {\bibfnamefont {R.~J.}\ \bibnamefont {Visser}}, \bibinfo {author} {\bibfnamefont {Z.}~\bibnamefont {Webb}}, \bibinfo {author} {\bibfnamefont {X.}~\bibnamefont {Zhan}}, \bibinfo {author} {\bibfnamefont {Y.}~\bibnamefont {Cohen}}, \bibinfo {author} {\bibfnamefont {P.}~\bibnamefont {Ronagh}}, \bibinfo {author} {\bibfnamefont {A.}~\bibnamefont {Ho}}, \bibinfo {author} {\bibfnamefont {R.~G.}\ \bibnamefont {Beausoleil}},\ and\ \bibinfo {author} {\bibfnamefont {J.~M.}\ \bibnamefont {Martinis}},\ }\href@noop {} {\bibinfo {title} {How to {Build} a {Quantum} {Supercomputer}: {Scaling} from {Hundreds} to {Millions} of {Qubits}}} (\bibinfo {year} {2025}),\ \Eprint {https://arxiv.org/abs/2411.10406} {arXiv:2411.10406} \BibitemShut {NoStop}%
\bibitem [{\citenamefont {Kosen}\ \emph {et~al.}(2024)\citenamefont {Kosen}, \citenamefont {Li}, \citenamefont {Rommel}, \citenamefont {Rehammar}, \citenamefont {Caputo}, \citenamefont {Grönberg}, \citenamefont {Fernández-Pendás}, \citenamefont {Kockum}, \citenamefont {Biznárová}, \citenamefont {Chen}, \citenamefont {Križan}, \citenamefont {Nylander}, \citenamefont {Osman}, \citenamefont {Roudsari}, \citenamefont {Shiri}, \citenamefont {Tancredi}, \citenamefont {Govenius},\ and\ \citenamefont {Bylander}}]{kosen_signal_2024}%
  \BibitemOpen
  \bibfield  {author} {\bibinfo {author} {\bibfnamefont {S.}~\bibnamefont {Kosen}}, \bibinfo {author} {\bibfnamefont {H.-X.}\ \bibnamefont {Li}}, \bibinfo {author} {\bibfnamefont {M.}~\bibnamefont {Rommel}}, \bibinfo {author} {\bibfnamefont {R.}~\bibnamefont {Rehammar}}, \bibinfo {author} {\bibfnamefont {M.}~\bibnamefont {Caputo}}, \bibinfo {author} {\bibfnamefont {L.}~\bibnamefont {Grönberg}}, \bibinfo {author} {\bibfnamefont {J.}~\bibnamefont {Fernández-Pendás}}, \bibinfo {author} {\bibfnamefont {A.~F.}\ \bibnamefont {Kockum}}, \bibinfo {author} {\bibfnamefont {J.}~\bibnamefont {Biznárová}}, \bibinfo {author} {\bibfnamefont {L.}~\bibnamefont {Chen}}, \bibinfo {author} {\bibfnamefont {C.}~\bibnamefont {Križan}}, \bibinfo {author} {\bibfnamefont {A.}~\bibnamefont {Nylander}}, \bibinfo {author} {\bibfnamefont {A.}~\bibnamefont {Osman}}, \bibinfo {author} {\bibfnamefont {A.~F.}\ \bibnamefont {Roudsari}}, \bibinfo {author} {\bibfnamefont {D.}~\bibnamefont {Shiri}}, \bibinfo {author} {\bibfnamefont
  {G.}~\bibnamefont {Tancredi}}, \bibinfo {author} {\bibfnamefont {J.}~\bibnamefont {Govenius}},\ and\ \bibinfo {author} {\bibfnamefont {J.}~\bibnamefont {Bylander}},\ }\href {https://doi.org/10.1103/PRXQuantum.5.030350} {\bibfield  {journal} {\bibinfo  {journal} {PRX Quantum}\ }\textbf {\bibinfo {volume} {5}},\ \bibinfo {pages} {030350} (\bibinfo {year} {2024})}\BibitemShut {NoStop}%
\bibitem [{\citenamefont {Strohm}\ \emph {et~al.}(2024)\citenamefont {Strohm}, \citenamefont {Wintersperger}, \citenamefont {Dommert}, \citenamefont {Basilewitsch}, \citenamefont {Reuber}, \citenamefont {Hoursanov}, \citenamefont {Ehmer}, \citenamefont {Vodola},\ and\ \citenamefont {Luber}}]{Strohm2024}%
  \BibitemOpen
  \bibfield  {author} {\bibinfo {author} {\bibfnamefont {T.}~\bibnamefont {Strohm}}, \bibinfo {author} {\bibfnamefont {K.}~\bibnamefont {Wintersperger}}, \bibinfo {author} {\bibfnamefont {F.}~\bibnamefont {Dommert}}, \bibinfo {author} {\bibfnamefont {D.}~\bibnamefont {Basilewitsch}}, \bibinfo {author} {\bibfnamefont {G.}~\bibnamefont {Reuber}}, \bibinfo {author} {\bibfnamefont {A.}~\bibnamefont {Hoursanov}}, \bibinfo {author} {\bibfnamefont {T.}~\bibnamefont {Ehmer}}, \bibinfo {author} {\bibfnamefont {D.}~\bibnamefont {Vodola}},\ and\ \bibinfo {author} {\bibfnamefont {S.}~\bibnamefont {Luber}},\ }\href@noop {} {\bibinfo {title} {{Ion-Based Quantum Computing Hardware: Performance and End-User Perspective}}} (\bibinfo {year} {2024}),\ \Eprint {https://arxiv.org/abs/2405.11450} {arXiv:2405.11450} \BibitemShut {NoStop}%
\bibitem [{\citenamefont {Wintersperger}\ \emph {et~al.}(2023)\citenamefont {Wintersperger}, \citenamefont {Dommert}, \citenamefont {Ehmer}, \citenamefont {Hoursanov}, \citenamefont {Klepsch}, \citenamefont {Mauerer}, \citenamefont {Reuber}, \citenamefont {Strohm}, \citenamefont {Yin},\ and\ \citenamefont {Luber}}]{Wintersperger2023}%
  \BibitemOpen
  \bibfield  {author} {\bibinfo {author} {\bibfnamefont {K.}~\bibnamefont {Wintersperger}}, \bibinfo {author} {\bibfnamefont {F.}~\bibnamefont {Dommert}}, \bibinfo {author} {\bibfnamefont {T.}~\bibnamefont {Ehmer}}, \bibinfo {author} {\bibfnamefont {A.}~\bibnamefont {Hoursanov}}, \bibinfo {author} {\bibfnamefont {J.}~\bibnamefont {Klepsch}}, \bibinfo {author} {\bibfnamefont {W.}~\bibnamefont {Mauerer}}, \bibinfo {author} {\bibfnamefont {G.}~\bibnamefont {Reuber}}, \bibinfo {author} {\bibfnamefont {T.}~\bibnamefont {Strohm}}, \bibinfo {author} {\bibfnamefont {M.}~\bibnamefont {Yin}},\ and\ \bibinfo {author} {\bibfnamefont {S.}~\bibnamefont {Luber}},\ }\href {https://doi.org/10.1140/epjqt/s40507-023-00190-1} {\bibfield  {journal} {\bibinfo  {journal} {EPJ Quantum Technology}\ }\textbf {\bibinfo {volume} {10}},\ \bibinfo {pages} {32} (\bibinfo {year} {2023})}\BibitemShut {NoStop}%
\bibitem [{\citenamefont {Monroe}\ and\ \citenamefont {Kim}(2013)}]{monroe_scaling_2013}%
  \BibitemOpen
  \bibfield  {author} {\bibinfo {author} {\bibfnamefont {C.}~\bibnamefont {Monroe}}\ and\ \bibinfo {author} {\bibfnamefont {J.}~\bibnamefont {Kim}},\ }\href {https://doi.org/10.1126/science.1231298} {\bibfield  {journal} {\bibinfo  {journal} {Science}\ }\textbf {\bibinfo {volume} {339}},\ \bibinfo {pages} {1164} (\bibinfo {year} {2013})}\BibitemShut {NoStop}%
\bibitem [{\citenamefont {Jiang}\ \emph {et~al.}(2007)\citenamefont {Jiang}, \citenamefont {Taylor}, \citenamefont {Sørensen},\ and\ \citenamefont {Lukin}}]{jiang_distributed_2007}%
  \BibitemOpen
  \bibfield  {author} {\bibinfo {author} {\bibfnamefont {L.}~\bibnamefont {Jiang}}, \bibinfo {author} {\bibfnamefont {J.~M.}\ \bibnamefont {Taylor}}, \bibinfo {author} {\bibfnamefont {A.~S.}\ \bibnamefont {Sørensen}},\ and\ \bibinfo {author} {\bibfnamefont {M.~D.}\ \bibnamefont {Lukin}},\ }\href {https://doi.org/10.1103/PhysRevA.76.062323} {\bibfield  {journal} {\bibinfo  {journal} {Physical Review A}\ }\textbf {\bibinfo {volume} {76}},\ \bibinfo {pages} {062323} (\bibinfo {year} {2007})}\BibitemShut {NoStop}%
\bibitem [{\citenamefont {Monroe}\ \emph {et~al.}(2014)\citenamefont {Monroe}, \citenamefont {Raussendorf}, \citenamefont {Ruthven}, \citenamefont {Brown}, \citenamefont {Maunz}, \citenamefont {Duan},\ and\ \citenamefont {Kim}}]{monroe_large-scale_2014}%
  \BibitemOpen
  \bibfield  {author} {\bibinfo {author} {\bibfnamefont {C.}~\bibnamefont {Monroe}}, \bibinfo {author} {\bibfnamefont {R.}~\bibnamefont {Raussendorf}}, \bibinfo {author} {\bibfnamefont {A.}~\bibnamefont {Ruthven}}, \bibinfo {author} {\bibfnamefont {K.~R.}\ \bibnamefont {Brown}}, \bibinfo {author} {\bibfnamefont {P.}~\bibnamefont {Maunz}}, \bibinfo {author} {\bibfnamefont {L.-M.}\ \bibnamefont {Duan}},\ and\ \bibinfo {author} {\bibfnamefont {J.}~\bibnamefont {Kim}},\ }\href {https://doi.org/10.1103/PhysRevA.89.022317} {\bibfield  {journal} {\bibinfo  {journal} {Physical Review A}\ }\textbf {\bibinfo {volume} {89}},\ \bibinfo {pages} {022317} (\bibinfo {year} {2014})}\BibitemShut {NoStop}%
\bibitem [{\citenamefont {Nickerson}\ \emph {et~al.}(2013)\citenamefont {Nickerson}, \citenamefont {Li},\ and\ \citenamefont {Benjamin}}]{nickerson_topological_2013}%
  \BibitemOpen
  \bibfield  {author} {\bibinfo {author} {\bibfnamefont {N.~H.}\ \bibnamefont {Nickerson}}, \bibinfo {author} {\bibfnamefont {Y.}~\bibnamefont {Li}},\ and\ \bibinfo {author} {\bibfnamefont {S.~C.}\ \bibnamefont {Benjamin}},\ }\href {https://doi.org/10.1038/ncomms2773} {\bibfield  {journal} {\bibinfo  {journal} {Nature Communications}\ }\textbf {\bibinfo {volume} {4}},\ \bibinfo {pages} {1756} (\bibinfo {year} {2013})}\BibitemShut {NoStop}%
\bibitem [{\citenamefont {Nickerson}\ \emph {et~al.}(2014)\citenamefont {Nickerson}, \citenamefont {Fitzsimons},\ and\ \citenamefont {Benjamin}}]{nickerson_freely_2014}%
  \BibitemOpen
  \bibfield  {author} {\bibinfo {author} {\bibfnamefont {N.~H.}\ \bibnamefont {Nickerson}}, \bibinfo {author} {\bibfnamefont {J.~F.}\ \bibnamefont {Fitzsimons}},\ and\ \bibinfo {author} {\bibfnamefont {S.~C.}\ \bibnamefont {Benjamin}},\ }\href {https://doi.org/10.1103/PhysRevX.4.041041} {\bibfield  {journal} {\bibinfo  {journal} {Physical Review X}\ }\textbf {\bibinfo {volume} {4}},\ \bibinfo {pages} {041041} (\bibinfo {year} {2014})}\BibitemShut {NoStop}%
\bibitem [{\citenamefont {Li}\ and\ \citenamefont {Benjamin}(2016)}]{li_hierarchical_2016}%
  \BibitemOpen
  \bibfield  {author} {\bibinfo {author} {\bibfnamefont {Y.}~\bibnamefont {Li}}\ and\ \bibinfo {author} {\bibfnamefont {S.~C.}\ \bibnamefont {Benjamin}},\ }\href {https://doi.org/10.1103/PhysRevA.94.042303} {\bibfield  {journal} {\bibinfo  {journal} {Physical Review A}\ }\textbf {\bibinfo {volume} {94}},\ \bibinfo {pages} {042303} (\bibinfo {year} {2016})}\BibitemShut {NoStop}%
\bibitem [{\citenamefont {Aghaee~Rad}\ \emph {et~al.}(2025)\citenamefont {Aghaee~Rad}, \citenamefont {Ainsworth}, \citenamefont {Alexander}, \citenamefont {Altieri}, \citenamefont {Askarani}, \citenamefont {Baby}, \citenamefont {Banchi}, \citenamefont {Baragiola}, \citenamefont {Bourassa}, \citenamefont {Chadwick}, \citenamefont {Charania}, \citenamefont {Chen}, \citenamefont {Collins}, \citenamefont {Contu}, \citenamefont {D’Arcy}, \citenamefont {Dauphinais}, \citenamefont {De~Prins}, \citenamefont {Deschenes}, \citenamefont {Di~Luch}, \citenamefont {Duque}, \citenamefont {Edke}, \citenamefont {Fayer}, \citenamefont {Ferracin}, \citenamefont {Ferretti}, \citenamefont {Gefaell}, \citenamefont {Glancy}, \citenamefont {González-Arciniegas}, \citenamefont {Grainge}, \citenamefont {Han}, \citenamefont {Hastrup}, \citenamefont {Helt}, \citenamefont {Hillmann}, \citenamefont {Hundal}, \citenamefont {Izumi}, \citenamefont {Jaeken}, \citenamefont {Jonas}, \citenamefont {Kocsis}, \citenamefont {Krasnokutska},
  \citenamefont {Larsen}, \citenamefont {Laskowski}, \citenamefont {Laudenbach}, \citenamefont {Lavoie}, \citenamefont {Li}, \citenamefont {Lomonte}, \citenamefont {Lopetegui}, \citenamefont {Luey}, \citenamefont {Lund}, \citenamefont {Ma}, \citenamefont {Madsen}, \citenamefont {Mahler}, \citenamefont {Mantilla~Calderón}, \citenamefont {Menotti}, \citenamefont {Miatto}, \citenamefont {Morrison}, \citenamefont {Nadkarni}, \citenamefont {Nakamura}, \citenamefont {Neuhaus}, \citenamefont {Niu}, \citenamefont {Noro}, \citenamefont {Papirov}, \citenamefont {Pesah}, \citenamefont {Phillips}, \citenamefont {Plick}, \citenamefont {Rogalsky}, \citenamefont {Rortais}, \citenamefont {Sabines-Chesterking}, \citenamefont {Safavi-Bayat}, \citenamefont {Sazhaev}, \citenamefont {Seymour}, \citenamefont {Rezaei~Shad}, \citenamefont {Silverman}, \citenamefont {Srinivasan}, \citenamefont {Stephan}, \citenamefont {Tang}, \citenamefont {Tasker}, \citenamefont {Teo}, \citenamefont {Then}, \citenamefont {Tremblay}, \citenamefont
  {Tzitrin}, \citenamefont {Vaidya}, \citenamefont {Vasmer}, \citenamefont {Vernon}, \citenamefont {Villalobos}, \citenamefont {Walshe}, \citenamefont {Weil}, \citenamefont {Xin}, \citenamefont {Yan}, \citenamefont {Yao}, \citenamefont {Zamani~Abnili},\ and\ \citenamefont {Zhang}}]{aghaee_rad_scaling_2025}%
  \BibitemOpen
  \bibfield  {author} {\bibinfo {author} {\bibfnamefont {H.}~\bibnamefont {Aghaee~Rad}}, \bibinfo {author} {\bibfnamefont {T.}~\bibnamefont {Ainsworth}}, \bibinfo {author} {\bibfnamefont {R.~N.}\ \bibnamefont {Alexander}}, \bibinfo {author} {\bibfnamefont {B.}~\bibnamefont {Altieri}}, \bibinfo {author} {\bibfnamefont {M.~F.}\ \bibnamefont {Askarani}}, \bibinfo {author} {\bibfnamefont {R.}~\bibnamefont {Baby}}, \bibinfo {author} {\bibfnamefont {L.}~\bibnamefont {Banchi}}, \bibinfo {author} {\bibfnamefont {B.~Q.}\ \bibnamefont {Baragiola}}, \bibinfo {author} {\bibfnamefont {J.~E.}\ \bibnamefont {Bourassa}}, \bibinfo {author} {\bibfnamefont {R.~S.}\ \bibnamefont {Chadwick}}, \bibinfo {author} {\bibfnamefont {I.}~\bibnamefont {Charania}}, \bibinfo {author} {\bibfnamefont {H.}~\bibnamefont {Chen}}, \bibinfo {author} {\bibfnamefont {M.~J.}\ \bibnamefont {Collins}}, \bibinfo {author} {\bibfnamefont {P.}~\bibnamefont {Contu}}, \bibinfo {author} {\bibfnamefont {N.}~\bibnamefont {D’Arcy}}, \bibinfo {author}
  {\bibfnamefont {G.}~\bibnamefont {Dauphinais}}, \bibinfo {author} {\bibfnamefont {R.}~\bibnamefont {De~Prins}}, \bibinfo {author} {\bibfnamefont {D.}~\bibnamefont {Deschenes}}, \bibinfo {author} {\bibfnamefont {I.}~\bibnamefont {Di~Luch}}, \bibinfo {author} {\bibfnamefont {S.}~\bibnamefont {Duque}}, \bibinfo {author} {\bibfnamefont {P.}~\bibnamefont {Edke}}, \bibinfo {author} {\bibfnamefont {S.~E.}\ \bibnamefont {Fayer}}, \bibinfo {author} {\bibfnamefont {S.}~\bibnamefont {Ferracin}}, \bibinfo {author} {\bibfnamefont {H.}~\bibnamefont {Ferretti}}, \bibinfo {author} {\bibfnamefont {J.}~\bibnamefont {Gefaell}}, \bibinfo {author} {\bibfnamefont {S.}~\bibnamefont {Glancy}}, \bibinfo {author} {\bibfnamefont {C.}~\bibnamefont {González-Arciniegas}}, \bibinfo {author} {\bibfnamefont {T.}~\bibnamefont {Grainge}}, \bibinfo {author} {\bibfnamefont {Z.}~\bibnamefont {Han}}, \bibinfo {author} {\bibfnamefont {J.}~\bibnamefont {Hastrup}}, \bibinfo {author} {\bibfnamefont {L.~G.}\ \bibnamefont {Helt}}, \bibinfo {author}
  {\bibfnamefont {T.}~\bibnamefont {Hillmann}}, \bibinfo {author} {\bibfnamefont {J.}~\bibnamefont {Hundal}}, \bibinfo {author} {\bibfnamefont {S.}~\bibnamefont {Izumi}}, \bibinfo {author} {\bibfnamefont {T.}~\bibnamefont {Jaeken}}, \bibinfo {author} {\bibfnamefont {M.}~\bibnamefont {Jonas}}, \bibinfo {author} {\bibfnamefont {S.}~\bibnamefont {Kocsis}}, \bibinfo {author} {\bibfnamefont {I.}~\bibnamefont {Krasnokutska}}, \bibinfo {author} {\bibfnamefont {M.~V.}\ \bibnamefont {Larsen}}, \bibinfo {author} {\bibfnamefont {P.}~\bibnamefont {Laskowski}}, \bibinfo {author} {\bibfnamefont {F.}~\bibnamefont {Laudenbach}}, \bibinfo {author} {\bibfnamefont {J.}~\bibnamefont {Lavoie}}, \bibinfo {author} {\bibfnamefont {M.}~\bibnamefont {Li}}, \bibinfo {author} {\bibfnamefont {E.}~\bibnamefont {Lomonte}}, \bibinfo {author} {\bibfnamefont {C.~E.}\ \bibnamefont {Lopetegui}}, \bibinfo {author} {\bibfnamefont {B.}~\bibnamefont {Luey}}, \bibinfo {author} {\bibfnamefont {A.~P.}\ \bibnamefont {Lund}}, \bibinfo {author}
  {\bibfnamefont {C.}~\bibnamefont {Ma}}, \bibinfo {author} {\bibfnamefont {L.~S.}\ \bibnamefont {Madsen}}, \bibinfo {author} {\bibfnamefont {D.~H.}\ \bibnamefont {Mahler}}, \bibinfo {author} {\bibfnamefont {L.}~\bibnamefont {Mantilla~Calderón}}, \bibinfo {author} {\bibfnamefont {M.}~\bibnamefont {Menotti}}, \bibinfo {author} {\bibfnamefont {F.~M.}\ \bibnamefont {Miatto}}, \bibinfo {author} {\bibfnamefont {B.}~\bibnamefont {Morrison}}, \bibinfo {author} {\bibfnamefont {P.~J.}\ \bibnamefont {Nadkarni}}, \bibinfo {author} {\bibfnamefont {T.}~\bibnamefont {Nakamura}}, \bibinfo {author} {\bibfnamefont {L.}~\bibnamefont {Neuhaus}}, \bibinfo {author} {\bibfnamefont {Z.}~\bibnamefont {Niu}}, \bibinfo {author} {\bibfnamefont {R.}~\bibnamefont {Noro}}, \bibinfo {author} {\bibfnamefont {K.}~\bibnamefont {Papirov}}, \bibinfo {author} {\bibfnamefont {A.}~\bibnamefont {Pesah}}, \bibinfo {author} {\bibfnamefont {D.~S.}\ \bibnamefont {Phillips}}, \bibinfo {author} {\bibfnamefont {W.~N.}\ \bibnamefont {Plick}}, \bibinfo
  {author} {\bibfnamefont {T.}~\bibnamefont {Rogalsky}}, \bibinfo {author} {\bibfnamefont {F.}~\bibnamefont {Rortais}}, \bibinfo {author} {\bibfnamefont {J.}~\bibnamefont {Sabines-Chesterking}}, \bibinfo {author} {\bibfnamefont {S.}~\bibnamefont {Safavi-Bayat}}, \bibinfo {author} {\bibfnamefont {E.}~\bibnamefont {Sazhaev}}, \bibinfo {author} {\bibfnamefont {M.}~\bibnamefont {Seymour}}, \bibinfo {author} {\bibfnamefont {K.}~\bibnamefont {Rezaei~Shad}}, \bibinfo {author} {\bibfnamefont {M.}~\bibnamefont {Silverman}}, \bibinfo {author} {\bibfnamefont {S.~A.}\ \bibnamefont {Srinivasan}}, \bibinfo {author} {\bibfnamefont {M.}~\bibnamefont {Stephan}}, \bibinfo {author} {\bibfnamefont {Q.~Y.}\ \bibnamefont {Tang}}, \bibinfo {author} {\bibfnamefont {J.~F.}\ \bibnamefont {Tasker}}, \bibinfo {author} {\bibfnamefont {Y.~S.}\ \bibnamefont {Teo}}, \bibinfo {author} {\bibfnamefont {R.~B.}\ \bibnamefont {Then}}, \bibinfo {author} {\bibfnamefont {J.~E.}\ \bibnamefont {Tremblay}}, \bibinfo {author} {\bibfnamefont
  {I.}~\bibnamefont {Tzitrin}}, \bibinfo {author} {\bibfnamefont {V.~D.}\ \bibnamefont {Vaidya}}, \bibinfo {author} {\bibfnamefont {M.}~\bibnamefont {Vasmer}}, \bibinfo {author} {\bibfnamefont {Z.}~\bibnamefont {Vernon}}, \bibinfo {author} {\bibfnamefont {L.~F. S. S.~M.}\ \bibnamefont {Villalobos}}, \bibinfo {author} {\bibfnamefont {B.~W.}\ \bibnamefont {Walshe}}, \bibinfo {author} {\bibfnamefont {R.}~\bibnamefont {Weil}}, \bibinfo {author} {\bibfnamefont {X.}~\bibnamefont {Xin}}, \bibinfo {author} {\bibfnamefont {X.}~\bibnamefont {Yan}}, \bibinfo {author} {\bibfnamefont {Y.}~\bibnamefont {Yao}}, \bibinfo {author} {\bibfnamefont {M.}~\bibnamefont {Zamani~Abnili}},\ and\ \bibinfo {author} {\bibfnamefont {Y.}~\bibnamefont {Zhang}},\ }\href {https://doi.org/10.1038/s41586-024-08406-9} {\bibfield  {journal} {\bibinfo  {journal} {Nature}\ }\textbf {\bibinfo {volume} {638}},\ \bibinfo {pages} {912} (\bibinfo {year} {2025})}\BibitemShut {NoStop}%
\bibitem [{\citenamefont {Ramette}\ \emph {et~al.}(2024)\citenamefont {Ramette}, \citenamefont {Sinclair}, \citenamefont {Breuckmann},\ and\ \citenamefont {Vuletić}}]{ramette_fault-tolerant_2024}%
  \BibitemOpen
  \bibfield  {author} {\bibinfo {author} {\bibfnamefont {J.}~\bibnamefont {Ramette}}, \bibinfo {author} {\bibfnamefont {J.}~\bibnamefont {Sinclair}}, \bibinfo {author} {\bibfnamefont {N.~P.}\ \bibnamefont {Breuckmann}},\ and\ \bibinfo {author} {\bibfnamefont {V.}~\bibnamefont {Vuletić}},\ }\href {https://doi.org/10.1038/s41534-024-00855-4} {\bibfield  {journal} {\bibinfo  {journal} {npj Quantum Information}\ }\textbf {\bibinfo {volume} {10}},\ \bibinfo {pages} {58} (\bibinfo {year} {2024})}\BibitemShut {NoStop}%
\bibitem [{\citenamefont {Gold}\ \emph {et~al.}(2021)\citenamefont {Gold}, \citenamefont {Paquette}, \citenamefont {Stockklauser}, \citenamefont {Reagor}, \citenamefont {Alam}, \citenamefont {Bestwick}, \citenamefont {Didier}, \citenamefont {Nersisyan}, \citenamefont {Oruc}, \citenamefont {Razavi}, \citenamefont {Scharmann}, \citenamefont {Sete}, \citenamefont {Sur}, \citenamefont {Venturelli}, \citenamefont {Winkleblack}, \citenamefont {Wudarski}, \citenamefont {Harburn},\ and\ \citenamefont {Rigetti}}]{gold_entanglement_2021}%
  \BibitemOpen
  \bibfield  {author} {\bibinfo {author} {\bibfnamefont {A.}~\bibnamefont {Gold}}, \bibinfo {author} {\bibfnamefont {J.~P.}\ \bibnamefont {Paquette}}, \bibinfo {author} {\bibfnamefont {A.}~\bibnamefont {Stockklauser}}, \bibinfo {author} {\bibfnamefont {M.~J.}\ \bibnamefont {Reagor}}, \bibinfo {author} {\bibfnamefont {M.~S.}\ \bibnamefont {Alam}}, \bibinfo {author} {\bibfnamefont {A.}~\bibnamefont {Bestwick}}, \bibinfo {author} {\bibfnamefont {N.}~\bibnamefont {Didier}}, \bibinfo {author} {\bibfnamefont {A.}~\bibnamefont {Nersisyan}}, \bibinfo {author} {\bibfnamefont {F.}~\bibnamefont {Oruc}}, \bibinfo {author} {\bibfnamefont {A.}~\bibnamefont {Razavi}}, \bibinfo {author} {\bibfnamefont {B.}~\bibnamefont {Scharmann}}, \bibinfo {author} {\bibfnamefont {E.~A.}\ \bibnamefont {Sete}}, \bibinfo {author} {\bibfnamefont {B.}~\bibnamefont {Sur}}, \bibinfo {author} {\bibfnamefont {D.}~\bibnamefont {Venturelli}}, \bibinfo {author} {\bibfnamefont {C.~J.}\ \bibnamefont {Winkleblack}}, \bibinfo {author} {\bibfnamefont
  {F.}~\bibnamefont {Wudarski}}, \bibinfo {author} {\bibfnamefont {M.}~\bibnamefont {Harburn}},\ and\ \bibinfo {author} {\bibfnamefont {C.}~\bibnamefont {Rigetti}},\ }\href {https://doi.org/10.1038/s41534-021-00484-1} {\bibfield  {journal} {\bibinfo  {journal} {npj Quantum Information}\ }\textbf {\bibinfo {volume} {7}},\ \bibinfo {pages} {142} (\bibinfo {year} {2021})}\BibitemShut {NoStop}%
\bibitem [{\citenamefont {Wu}\ \emph {et~al.}(2024)\citenamefont {Wu}, \citenamefont {Yan}, \citenamefont {Andersson}, \citenamefont {Anferov}, \citenamefont {Chou}, \citenamefont {Conner}, \citenamefont {Grebel}, \citenamefont {Joshi}, \citenamefont {Li}, \citenamefont {Miller}, \citenamefont {Povey}, \citenamefont {Qiao},\ and\ \citenamefont {Cleland}}]{wu_modular_2024}%
  \BibitemOpen
  \bibfield  {author} {\bibinfo {author} {\bibfnamefont {X.}~\bibnamefont {Wu}}, \bibinfo {author} {\bibfnamefont {H.}~\bibnamefont {Yan}}, \bibinfo {author} {\bibfnamefont {G.}~\bibnamefont {Andersson}}, \bibinfo {author} {\bibfnamefont {A.}~\bibnamefont {Anferov}}, \bibinfo {author} {\bibfnamefont {M.-H.}\ \bibnamefont {Chou}}, \bibinfo {author} {\bibfnamefont {C.~R.}\ \bibnamefont {Conner}}, \bibinfo {author} {\bibfnamefont {J.}~\bibnamefont {Grebel}}, \bibinfo {author} {\bibfnamefont {Y.~J.}\ \bibnamefont {Joshi}}, \bibinfo {author} {\bibfnamefont {S.}~\bibnamefont {Li}}, \bibinfo {author} {\bibfnamefont {J.~M.}\ \bibnamefont {Miller}}, \bibinfo {author} {\bibfnamefont {R.~G.}\ \bibnamefont {Povey}}, \bibinfo {author} {\bibfnamefont {H.}~\bibnamefont {Qiao}},\ and\ \bibinfo {author} {\bibfnamefont {A.~N.}\ \bibnamefont {Cleland}},\ }\href {https://doi.org/10.1103/PhysRevX.14.041030} {\bibfield  {journal} {\bibinfo  {journal} {Physical Review X}\ }\textbf {\bibinfo {volume} {14}},\ \bibinfo {pages}
  {041030} (\bibinfo {year} {2024})}\BibitemShut {NoStop}%
\bibitem [{\citenamefont {Mollenhauer}\ \emph {et~al.}(2024)\citenamefont {Mollenhauer}, \citenamefont {Irfan}, \citenamefont {Cao}, \citenamefont {Mandal},\ and\ \citenamefont {Pfaff}}]{mollenhauer_high-efficiency_2024}%
  \BibitemOpen
  \bibfield  {author} {\bibinfo {author} {\bibfnamefont {M.}~\bibnamefont {Mollenhauer}}, \bibinfo {author} {\bibfnamefont {A.}~\bibnamefont {Irfan}}, \bibinfo {author} {\bibfnamefont {X.}~\bibnamefont {Cao}}, \bibinfo {author} {\bibfnamefont {S.}~\bibnamefont {Mandal}},\ and\ \bibinfo {author} {\bibfnamefont {W.}~\bibnamefont {Pfaff}},\ }\href@noop {} {\bibinfo {title} {A high-efficiency plug-and-play superconducting qubit network}} (\bibinfo {year} {2024}),\ \Eprint {https://arxiv.org/abs/2407.16743} {arXiv:2407.16743} \BibitemShut {NoStop}%
\bibitem [{\citenamefont {Roch}\ \emph {et~al.}(2014)\citenamefont {Roch}, \citenamefont {Schwartz}, \citenamefont {Motzoi}, \citenamefont {Macklin}, \citenamefont {Vijay}, \citenamefont {Eddins}, \citenamefont {Korotkov}, \citenamefont {Whaley}, \citenamefont {Sarovar},\ and\ \citenamefont {Siddiqi}}]{roch_observation_2014}%
  \BibitemOpen
  \bibfield  {author} {\bibinfo {author} {\bibfnamefont {N.}~\bibnamefont {Roch}}, \bibinfo {author} {\bibfnamefont {M.~E.}\ \bibnamefont {Schwartz}}, \bibinfo {author} {\bibfnamefont {F.}~\bibnamefont {Motzoi}}, \bibinfo {author} {\bibfnamefont {C.}~\bibnamefont {Macklin}}, \bibinfo {author} {\bibfnamefont {R.}~\bibnamefont {Vijay}}, \bibinfo {author} {\bibfnamefont {A.~W.}\ \bibnamefont {Eddins}}, \bibinfo {author} {\bibfnamefont {A.~N.}\ \bibnamefont {Korotkov}}, \bibinfo {author} {\bibfnamefont {K.~B.}\ \bibnamefont {Whaley}}, \bibinfo {author} {\bibfnamefont {M.}~\bibnamefont {Sarovar}},\ and\ \bibinfo {author} {\bibfnamefont {I.}~\bibnamefont {Siddiqi}},\ }\href {https://doi.org/10.1103/PhysRevLett.112.170501} {\bibfield  {journal} {\bibinfo  {journal} {Physical Review Letters}\ }\textbf {\bibinfo {volume} {112}},\ \bibinfo {pages} {170501} (\bibinfo {year} {2014})}\BibitemShut {NoStop}%
\bibitem [{\citenamefont {Hensen}\ \emph {et~al.}(2015)\citenamefont {Hensen}, \citenamefont {Bernien}, \citenamefont {Dréau}, \citenamefont {Reiserer}, \citenamefont {Kalb}, \citenamefont {Blok}, \citenamefont {Ruitenberg}, \citenamefont {Vermeulen}, \citenamefont {Schouten}, \citenamefont {Abellán}, \citenamefont {Amaya}, \citenamefont {Pruneri}, \citenamefont {Mitchell}, \citenamefont {Markham}, \citenamefont {Twitchen}, \citenamefont {Elkouss}, \citenamefont {Wehner}, \citenamefont {Taminiau},\ and\ \citenamefont {Hanson}}]{hensen_loophole-free_2015}%
  \BibitemOpen
  \bibfield  {author} {\bibinfo {author} {\bibfnamefont {B.}~\bibnamefont {Hensen}}, \bibinfo {author} {\bibfnamefont {H.}~\bibnamefont {Bernien}}, \bibinfo {author} {\bibfnamefont {A.~E.}\ \bibnamefont {Dréau}}, \bibinfo {author} {\bibfnamefont {A.}~\bibnamefont {Reiserer}}, \bibinfo {author} {\bibfnamefont {N.}~\bibnamefont {Kalb}}, \bibinfo {author} {\bibfnamefont {M.~S.}\ \bibnamefont {Blok}}, \bibinfo {author} {\bibfnamefont {J.}~\bibnamefont {Ruitenberg}}, \bibinfo {author} {\bibfnamefont {R.~F.~L.}\ \bibnamefont {Vermeulen}}, \bibinfo {author} {\bibfnamefont {R.~N.}\ \bibnamefont {Schouten}}, \bibinfo {author} {\bibfnamefont {C.}~\bibnamefont {Abellán}}, \bibinfo {author} {\bibfnamefont {W.}~\bibnamefont {Amaya}}, \bibinfo {author} {\bibfnamefont {V.}~\bibnamefont {Pruneri}}, \bibinfo {author} {\bibfnamefont {M.~W.}\ \bibnamefont {Mitchell}}, \bibinfo {author} {\bibfnamefont {M.}~\bibnamefont {Markham}}, \bibinfo {author} {\bibfnamefont {D.~J.}\ \bibnamefont {Twitchen}}, \bibinfo {author}
  {\bibfnamefont {D.}~\bibnamefont {Elkouss}}, \bibinfo {author} {\bibfnamefont {S.}~\bibnamefont {Wehner}}, \bibinfo {author} {\bibfnamefont {T.~H.}\ \bibnamefont {Taminiau}},\ and\ \bibinfo {author} {\bibfnamefont {R.}~\bibnamefont {Hanson}},\ }\href {https://doi.org/10.1038/nature15759} {\bibfield  {journal} {\bibinfo  {journal} {Nature}\ }\textbf {\bibinfo {volume} {526}},\ \bibinfo {pages} {682} (\bibinfo {year} {2015})}\BibitemShut {NoStop}%
\bibitem [{\citenamefont {Dickel}\ \emph {et~al.}(2018)\citenamefont {Dickel}, \citenamefont {Wesdorp}, \citenamefont {Langford}, \citenamefont {Peiter}, \citenamefont {Sagastizabal}, \citenamefont {Bruno}, \citenamefont {Criger}, \citenamefont {Motzoi},\ and\ \citenamefont {DiCarlo}}]{dickel_chip_2018}%
  \BibitemOpen
  \bibfield  {author} {\bibinfo {author} {\bibfnamefont {C.}~\bibnamefont {Dickel}}, \bibinfo {author} {\bibfnamefont {J.~J.}\ \bibnamefont {Wesdorp}}, \bibinfo {author} {\bibfnamefont {N.~K.}\ \bibnamefont {Langford}}, \bibinfo {author} {\bibfnamefont {S.}~\bibnamefont {Peiter}}, \bibinfo {author} {\bibfnamefont {R.}~\bibnamefont {Sagastizabal}}, \bibinfo {author} {\bibfnamefont {A.}~\bibnamefont {Bruno}}, \bibinfo {author} {\bibfnamefont {B.}~\bibnamefont {Criger}}, \bibinfo {author} {\bibfnamefont {F.}~\bibnamefont {Motzoi}},\ and\ \bibinfo {author} {\bibfnamefont {L.}~\bibnamefont {DiCarlo}},\ }\href {https://doi.org/10.1103/PhysRevB.97.064508} {\bibfield  {journal} {\bibinfo  {journal} {Physical Review B}\ }\textbf {\bibinfo {volume} {97}},\ \bibinfo {pages} {064508} (\bibinfo {year} {2018})}\BibitemShut {NoStop}%
\bibitem [{\citenamefont {Zhong}\ \emph {et~al.}(2021)\citenamefont {Zhong}, \citenamefont {Chang}, \citenamefont {Bienfait}, \citenamefont {Dumur}, \citenamefont {Chou}, \citenamefont {Conner}, \citenamefont {Grebel}, \citenamefont {Povey}, \citenamefont {Yan}, \citenamefont {Schuster},\ and\ \citenamefont {Cleland}}]{zhong_deterministic_2021}%
  \BibitemOpen
  \bibfield  {author} {\bibinfo {author} {\bibfnamefont {Y.}~\bibnamefont {Zhong}}, \bibinfo {author} {\bibfnamefont {H.-S.}\ \bibnamefont {Chang}}, \bibinfo {author} {\bibfnamefont {A.}~\bibnamefont {Bienfait}}, \bibinfo {author} {\bibfnamefont {E.}~\bibnamefont {Dumur}}, \bibinfo {author} {\bibfnamefont {M.-H.}\ \bibnamefont {Chou}}, \bibinfo {author} {\bibfnamefont {C.~R.}\ \bibnamefont {Conner}}, \bibinfo {author} {\bibfnamefont {J.}~\bibnamefont {Grebel}}, \bibinfo {author} {\bibfnamefont {R.~G.}\ \bibnamefont {Povey}}, \bibinfo {author} {\bibfnamefont {H.}~\bibnamefont {Yan}}, \bibinfo {author} {\bibfnamefont {D.~I.}\ \bibnamefont {Schuster}},\ and\ \bibinfo {author} {\bibfnamefont {A.~N.}\ \bibnamefont {Cleland}},\ }\href {https://doi.org/10.1038/s41586-021-03288-7} {\bibfield  {journal} {\bibinfo  {journal} {Nature}\ }\textbf {\bibinfo {volume} {590}},\ \bibinfo {pages} {571} (\bibinfo {year} {2021})}\BibitemShut {NoStop}%
\bibitem [{\citenamefont {Burkhart}\ \emph {et~al.}(2021)\citenamefont {Burkhart}, \citenamefont {Teoh}, \citenamefont {Zhang}, \citenamefont {Axline}, \citenamefont {Frunzio}, \citenamefont {Devoret}, \citenamefont {Jiang}, \citenamefont {Girvin},\ and\ \citenamefont {Schoelkopf}}]{burkhart_error_2021}%
  \BibitemOpen
  \bibfield  {author} {\bibinfo {author} {\bibfnamefont {L.~D.}\ \bibnamefont {Burkhart}}, \bibinfo {author} {\bibfnamefont {J.~D.}\ \bibnamefont {Teoh}}, \bibinfo {author} {\bibfnamefont {Y.}~\bibnamefont {Zhang}}, \bibinfo {author} {\bibfnamefont {C.~J.}\ \bibnamefont {Axline}}, \bibinfo {author} {\bibfnamefont {L.}~\bibnamefont {Frunzio}}, \bibinfo {author} {\bibfnamefont {M.}~\bibnamefont {Devoret}}, \bibinfo {author} {\bibfnamefont {L.}~\bibnamefont {Jiang}}, \bibinfo {author} {\bibfnamefont {S.}~\bibnamefont {Girvin}},\ and\ \bibinfo {author} {\bibfnamefont {R.}~\bibnamefont {Schoelkopf}},\ }\href {https://doi.org/10.1103/PRXQuantum.2.030321} {\bibfield  {journal} {\bibinfo  {journal} {PRX Quantum}\ }\textbf {\bibinfo {volume} {2}},\ \bibinfo {pages} {030321} (\bibinfo {year} {2021})}\BibitemShut {NoStop}%
\bibitem [{\citenamefont {Main}\ \emph {et~al.}(2025)\citenamefont {Main}, \citenamefont {Drmota}, \citenamefont {Nadlinger}, \citenamefont {Ainley}, \citenamefont {Agrawal}, \citenamefont {Nichol}, \citenamefont {Srinivas}, \citenamefont {Araneda},\ and\ \citenamefont {Lucas}}]{main_distributed_2025}%
  \BibitemOpen
  \bibfield  {author} {\bibinfo {author} {\bibfnamefont {D.}~\bibnamefont {Main}}, \bibinfo {author} {\bibfnamefont {P.}~\bibnamefont {Drmota}}, \bibinfo {author} {\bibfnamefont {D.~P.}\ \bibnamefont {Nadlinger}}, \bibinfo {author} {\bibfnamefont {E.~M.}\ \bibnamefont {Ainley}}, \bibinfo {author} {\bibfnamefont {A.}~\bibnamefont {Agrawal}}, \bibinfo {author} {\bibfnamefont {B.~C.}\ \bibnamefont {Nichol}}, \bibinfo {author} {\bibfnamefont {R.}~\bibnamefont {Srinivas}}, \bibinfo {author} {\bibfnamefont {G.}~\bibnamefont {Araneda}},\ and\ \bibinfo {author} {\bibfnamefont {D.~M.}\ \bibnamefont {Lucas}},\ }\href {https://doi.org/10.1038/s41586-024-08404-x} {\bibfield  {journal} {\bibinfo  {journal} {Nature}\ }\textbf {\bibinfo {volume} {638}},\ \bibinfo {pages} {383} (\bibinfo {year} {2025})}\BibitemShut {NoStop}%
\bibitem [{\citenamefont {Almanakly}\ \emph {et~al.}(2025)\citenamefont {Almanakly}, \citenamefont {Yankelevich}, \citenamefont {Hays}, \citenamefont {Kannan}, \citenamefont {Assouly}, \citenamefont {Greene}, \citenamefont {Gingras}, \citenamefont {Niedzielski}, \citenamefont {Stickler}, \citenamefont {Schwartz}, \citenamefont {Serniak}, \citenamefont {Wang}, \citenamefont {Orlando}, \citenamefont {Gustavsson}, \citenamefont {Grover},\ and\ \citenamefont {Oliver}}]{almanakly_deterministic_2025}%
  \BibitemOpen
  \bibfield  {author} {\bibinfo {author} {\bibfnamefont {A.}~\bibnamefont {Almanakly}}, \bibinfo {author} {\bibfnamefont {B.}~\bibnamefont {Yankelevich}}, \bibinfo {author} {\bibfnamefont {M.}~\bibnamefont {Hays}}, \bibinfo {author} {\bibfnamefont {B.}~\bibnamefont {Kannan}}, \bibinfo {author} {\bibfnamefont {R.}~\bibnamefont {Assouly}}, \bibinfo {author} {\bibfnamefont {A.}~\bibnamefont {Greene}}, \bibinfo {author} {\bibfnamefont {M.}~\bibnamefont {Gingras}}, \bibinfo {author} {\bibfnamefont {B.~M.}\ \bibnamefont {Niedzielski}}, \bibinfo {author} {\bibfnamefont {H.}~\bibnamefont {Stickler}}, \bibinfo {author} {\bibfnamefont {M.~E.}\ \bibnamefont {Schwartz}}, \bibinfo {author} {\bibfnamefont {K.}~\bibnamefont {Serniak}}, \bibinfo {author} {\bibfnamefont {J.~I.-j.}\ \bibnamefont {Wang}}, \bibinfo {author} {\bibfnamefont {T.~P.}\ \bibnamefont {Orlando}}, \bibinfo {author} {\bibfnamefont {S.}~\bibnamefont {Gustavsson}}, \bibinfo {author} {\bibfnamefont {J.~A.}\ \bibnamefont {Grover}},\ and\ \bibinfo {author}
  {\bibfnamefont {W.~D.}\ \bibnamefont {Oliver}},\ }\href {https://doi.org/10.1038/s41567-025-02811-1} {\bibfield  {journal} {\bibinfo  {journal} {Nature Physics}\ }\textbf {\bibinfo {volume} {21}},\ \bibinfo {pages} {825} (\bibinfo {year} {2025})}\BibitemShut {NoStop}%
\bibitem [{\citenamefont {Horsman}\ \emph {et~al.}(2012)\citenamefont {Horsman}, \citenamefont {Fowler}, \citenamefont {Devitt},\ and\ \citenamefont {Meter}}]{horsman_surface_2012}%
  \BibitemOpen
  \bibfield  {author} {\bibinfo {author} {\bibfnamefont {D.}~\bibnamefont {Horsman}}, \bibinfo {author} {\bibfnamefont {A.~G.}\ \bibnamefont {Fowler}}, \bibinfo {author} {\bibfnamefont {S.}~\bibnamefont {Devitt}},\ and\ \bibinfo {author} {\bibfnamefont {R.~V.}\ \bibnamefont {Meter}},\ }\href {https://doi.org/10.1088/1367-2630/14/12/123011} {\bibfield  {journal} {\bibinfo  {journal} {New Journal of Physics}\ }\textbf {\bibinfo {volume} {14}},\ \bibinfo {pages} {123011} (\bibinfo {year} {2012})}\BibitemShut {NoStop}%
\bibitem [{\citenamefont {Litinski}(2019)}]{litinski_game_2019}%
  \BibitemOpen
  \bibfield  {author} {\bibinfo {author} {\bibfnamefont {D.}~\bibnamefont {Litinski}},\ }\href {https://doi.org/10.22331/q-2019-03-05-128} {\bibfield  {journal} {\bibinfo  {journal} {Quantum}\ }\textbf {\bibinfo {volume} {3}},\ \bibinfo {pages} {128} (\bibinfo {year} {2019})}\BibitemShut {NoStop}%
\bibitem [{\citenamefont {Kitaev}(2003)}]{kitaev_fault-tolerant_2003}%
  \BibitemOpen
  \bibfield  {author} {\bibinfo {author} {\bibfnamefont {A.~{\relax Yu}.}\ \bibnamefont {Kitaev}},\ }\href {https://doi.org/10.1016/S0003-4916(02)00018-0} {\bibfield  {journal} {\bibinfo  {journal} {Annals of Physics}\ }\textbf {\bibinfo {volume} {303}},\ \bibinfo {pages} {2} (\bibinfo {year} {2003})}\BibitemShut {NoStop}%
\bibitem [{\citenamefont {Dennis}\ \emph {et~al.}(2002)\citenamefont {Dennis}, \citenamefont {Kitaev}, \citenamefont {Landahl},\ and\ \citenamefont {Preskill}}]{dennis_topological_2002}%
  \BibitemOpen
  \bibfield  {author} {\bibinfo {author} {\bibfnamefont {E.}~\bibnamefont {Dennis}}, \bibinfo {author} {\bibfnamefont {A.}~\bibnamefont {Kitaev}}, \bibinfo {author} {\bibfnamefont {A.}~\bibnamefont {Landahl}},\ and\ \bibinfo {author} {\bibfnamefont {J.}~\bibnamefont {Preskill}},\ }\href {https://doi.org/10.1063/1.1499754} {\bibfield  {journal} {\bibinfo  {journal} {Journal of Mathematical Physics}\ }\textbf {\bibinfo {volume} {43}},\ \bibinfo {pages} {4452} (\bibinfo {year} {2002})}\BibitemShut {NoStop}%
\bibitem [{\citenamefont {Bennett}\ \emph {et~al.}(1996)\citenamefont {Bennett}, \citenamefont {Bernstein}, \citenamefont {Popescu},\ and\ \citenamefont {Schumacher}}]{bennett_concentrating_1996}%
  \BibitemOpen
  \bibfield  {author} {\bibinfo {author} {\bibfnamefont {C.~H.}\ \bibnamefont {Bennett}}, \bibinfo {author} {\bibfnamefont {H.~J.}\ \bibnamefont {Bernstein}}, \bibinfo {author} {\bibfnamefont {S.}~\bibnamefont {Popescu}},\ and\ \bibinfo {author} {\bibfnamefont {B.}~\bibnamefont {Schumacher}},\ }\href {https://doi.org/10.1103/PhysRevA.53.2046} {\bibfield  {journal} {\bibinfo  {journal} {Physical Review A}\ }\textbf {\bibinfo {volume} {53}},\ \bibinfo {pages} {2046} (\bibinfo {year} {1996})}\BibitemShut {NoStop}%
\bibitem [{\citenamefont {Shalby}\ \emph {et~al.}(2025)\citenamefont {Shalby}, \citenamefont {Wang}, \citenamefont {Sedov},\ and\ \citenamefont {Pryadko}}]{shalby_optimized_2025}%
  \BibitemOpen
  \bibfield  {author} {\bibinfo {author} {\bibfnamefont {M.~A.}\ \bibnamefont {Shalby}}, \bibinfo {author} {\bibfnamefont {R.}~\bibnamefont {Wang}}, \bibinfo {author} {\bibfnamefont {D.}~\bibnamefont {Sedov}},\ and\ \bibinfo {author} {\bibfnamefont {L.~P.}\ \bibnamefont {Pryadko}},\ }\href@noop {} {\bibinfo {title} {Optimized noise-resilient surface code teleportation interfaces}} (\bibinfo {year} {2025}),\ \Eprint {https://arxiv.org/abs/2503.04968} {arXiv:2503.04968} \BibitemShut {NoStop}%
\bibitem [{\citenamefont {Jacinto}\ \emph {et~al.}(2025)\citenamefont {Jacinto}, \citenamefont {Gouzien},\ and\ \citenamefont {Sangouard}}]{jacinto_network_2025}%
  \BibitemOpen
  \bibfield  {author} {\bibinfo {author} {\bibfnamefont {H.}~\bibnamefont {Jacinto}}, \bibinfo {author} {\bibfnamefont {E.}~\bibnamefont {Gouzien}},\ and\ \bibinfo {author} {\bibfnamefont {N.}~\bibnamefont {Sangouard}},\ }\href@noop {} {\bibinfo {title} {Network {Requirements} for {Distributed} {Quantum} {Computation}}} (\bibinfo {year} {2025}),\ \Eprint {https://arxiv.org/abs/2504.08891} {arXiv:2504.08891} \BibitemShut {NoStop}%
\bibitem [{\citenamefont {Magnard}\ \emph {et~al.}(2020)\citenamefont {Magnard}, \citenamefont {Storz}, \citenamefont {Kurpiers}, \citenamefont {Schär}, \citenamefont {Marxer}, \citenamefont {Lütolf}, \citenamefont {Walter}, \citenamefont {Besse}, \citenamefont {Gabureac}, \citenamefont {Reuer}, \citenamefont {Akin}, \citenamefont {Royer}, \citenamefont {Blais},\ and\ \citenamefont {Wallraff}}]{magnard_microwave_2020}%
  \BibitemOpen
  \bibfield  {author} {\bibinfo {author} {\bibfnamefont {P.}~\bibnamefont {Magnard}}, \bibinfo {author} {\bibfnamefont {S.}~\bibnamefont {Storz}}, \bibinfo {author} {\bibfnamefont {P.}~\bibnamefont {Kurpiers}}, \bibinfo {author} {\bibfnamefont {J.}~\bibnamefont {Schär}}, \bibinfo {author} {\bibfnamefont {F.}~\bibnamefont {Marxer}}, \bibinfo {author} {\bibfnamefont {J.}~\bibnamefont {Lütolf}}, \bibinfo {author} {\bibfnamefont {T.}~\bibnamefont {Walter}}, \bibinfo {author} {\bibfnamefont {J.-C.}\ \bibnamefont {Besse}}, \bibinfo {author} {\bibfnamefont {M.}~\bibnamefont {Gabureac}}, \bibinfo {author} {\bibfnamefont {K.}~\bibnamefont {Reuer}}, \bibinfo {author} {\bibfnamefont {A.}~\bibnamefont {Akin}}, \bibinfo {author} {\bibfnamefont {B.}~\bibnamefont {Royer}}, \bibinfo {author} {\bibfnamefont {A.}~\bibnamefont {Blais}},\ and\ \bibinfo {author} {\bibfnamefont {A.}~\bibnamefont {Wallraff}},\ }\href {https://doi.org/10.1103/PhysRevLett.125.260502} {\bibfield  {journal} {\bibinfo  {journal} {Physical Review
  Letters}\ }\textbf {\bibinfo {volume} {125}},\ \bibinfo {pages} {260502} (\bibinfo {year} {2020})}\BibitemShut {NoStop}%
\bibitem [{\citenamefont {Stephenson}\ \emph {et~al.}(2020)\citenamefont {Stephenson}, \citenamefont {Nadlinger}, \citenamefont {Nichol}, \citenamefont {An}, \citenamefont {Drmota}, \citenamefont {Ballance}, \citenamefont {Thirumalai}, \citenamefont {Goodwin}, \citenamefont {Lucas},\ and\ \citenamefont {Ballance}}]{stephenson_high-rate_2020}%
  \BibitemOpen
  \bibfield  {author} {\bibinfo {author} {\bibfnamefont {L.~J.}\ \bibnamefont {Stephenson}}, \bibinfo {author} {\bibfnamefont {D.~P.}\ \bibnamefont {Nadlinger}}, \bibinfo {author} {\bibfnamefont {B.~C.}\ \bibnamefont {Nichol}}, \bibinfo {author} {\bibfnamefont {S.}~\bibnamefont {An}}, \bibinfo {author} {\bibfnamefont {P.}~\bibnamefont {Drmota}}, \bibinfo {author} {\bibfnamefont {T.~G.}\ \bibnamefont {Ballance}}, \bibinfo {author} {\bibfnamefont {K.}~\bibnamefont {Thirumalai}}, \bibinfo {author} {\bibfnamefont {J.~F.}\ \bibnamefont {Goodwin}}, \bibinfo {author} {\bibfnamefont {D.~M.}\ \bibnamefont {Lucas}},\ and\ \bibinfo {author} {\bibfnamefont {C.~J.}\ \bibnamefont {Ballance}},\ }\href {https://doi.org/10.1103/PhysRevLett.124.110501} {\bibfield  {journal} {\bibinfo  {journal} {Physical Review Letters}\ }\textbf {\bibinfo {volume} {124}},\ \bibinfo {pages} {110501} (\bibinfo {year} {2020})}\BibitemShut {NoStop}%
\bibitem [{\citenamefont {Heya}\ \emph {et~al.}(2025)\citenamefont {Heya}, \citenamefont {Phung}, \citenamefont {Malekakhlagh}, \citenamefont {Steiner}, \citenamefont {Turchetti}, \citenamefont {Shanks}, \citenamefont {Mamin}, \citenamefont {Lu}, \citenamefont {Kandel}, \citenamefont {Sundaresan},\ and\ \citenamefont {Orcutt}}]{heya_randomized_2025}%
  \BibitemOpen
  \bibfield  {author} {\bibinfo {author} {\bibfnamefont {K.}~\bibnamefont {Heya}}, \bibinfo {author} {\bibfnamefont {T.}~\bibnamefont {Phung}}, \bibinfo {author} {\bibfnamefont {M.}~\bibnamefont {Malekakhlagh}}, \bibinfo {author} {\bibfnamefont {R.}~\bibnamefont {Steiner}}, \bibinfo {author} {\bibfnamefont {M.}~\bibnamefont {Turchetti}}, \bibinfo {author} {\bibfnamefont {W.}~\bibnamefont {Shanks}}, \bibinfo {author} {\bibfnamefont {J.}~\bibnamefont {Mamin}}, \bibinfo {author} {\bibfnamefont {W.-S.}\ \bibnamefont {Lu}}, \bibinfo {author} {\bibfnamefont {Y.~P.}\ \bibnamefont {Kandel}}, \bibinfo {author} {\bibfnamefont {N.}~\bibnamefont {Sundaresan}},\ and\ \bibinfo {author} {\bibfnamefont {J.}~\bibnamefont {Orcutt}},\ }\href@noop {} {\bibinfo {title} {Randomized benchmarking of a high-fidelity remote {CNOT} gate over a meter-scale microwave interconnect}} (\bibinfo {year} {2025}),\ \Eprint {https://arxiv.org/abs/2502.15034} {arXiv:2502.15034} \BibitemShut {NoStop}%
\bibitem [{\citenamefont {Browne}\ and\ \citenamefont {Rudolph}(2005)}]{browne_resource-efficient_2005}%
  \BibitemOpen
  \bibfield  {author} {\bibinfo {author} {\bibfnamefont {D.~E.}\ \bibnamefont {Browne}}\ and\ \bibinfo {author} {\bibfnamefont {T.}~\bibnamefont {Rudolph}},\ }\href {https://doi.org/10.1103/PhysRevLett.95.010501} {\bibfield  {journal} {\bibinfo  {journal} {Physical Review Letters}\ }\textbf {\bibinfo {volume} {95}},\ \bibinfo {pages} {010501} (\bibinfo {year} {2005})}\BibitemShut {NoStop}%
\bibitem [{\citenamefont {Bartolucci}\ \emph {et~al.}(2023)\citenamefont {Bartolucci}, \citenamefont {Birchall}, \citenamefont {Bombín}, \citenamefont {Cable}, \citenamefont {Dawson}, \citenamefont {Gimeno-Segovia}, \citenamefont {Johnston}, \citenamefont {Kieling}, \citenamefont {Nickerson}, \citenamefont {Pant}, \citenamefont {Pastawski}, \citenamefont {Rudolph},\ and\ \citenamefont {Sparrow}}]{bartolucci_fusion_2023}%
  \BibitemOpen
  \bibfield  {author} {\bibinfo {author} {\bibfnamefont {S.}~\bibnamefont {Bartolucci}}, \bibinfo {author} {\bibfnamefont {P.}~\bibnamefont {Birchall}}, \bibinfo {author} {\bibfnamefont {H.}~\bibnamefont {Bombín}}, \bibinfo {author} {\bibfnamefont {H.}~\bibnamefont {Cable}}, \bibinfo {author} {\bibfnamefont {C.}~\bibnamefont {Dawson}}, \bibinfo {author} {\bibfnamefont {M.}~\bibnamefont {Gimeno-Segovia}}, \bibinfo {author} {\bibfnamefont {E.}~\bibnamefont {Johnston}}, \bibinfo {author} {\bibfnamefont {K.}~\bibnamefont {Kieling}}, \bibinfo {author} {\bibfnamefont {N.}~\bibnamefont {Nickerson}}, \bibinfo {author} {\bibfnamefont {M.}~\bibnamefont {Pant}}, \bibinfo {author} {\bibfnamefont {F.}~\bibnamefont {Pastawski}}, \bibinfo {author} {\bibfnamefont {T.}~\bibnamefont {Rudolph}},\ and\ \bibinfo {author} {\bibfnamefont {C.}~\bibnamefont {Sparrow}},\ }\href {https://doi.org/10.1038/s41467-023-36493-1} {\bibfield  {journal} {\bibinfo  {journal} {Nature Communications}\ }\textbf {\bibinfo {volume} {14}},\ \bibinfo
  {pages} {912} (\bibinfo {year} {2023})}\BibitemShut {NoStop}%
\bibitem [{\citenamefont {Verstraete}\ and\ \citenamefont {Cirac}(2004)}]{verstraete_valence_2004}%
  \BibitemOpen
  \bibfield  {author} {\bibinfo {author} {\bibfnamefont {F.}~\bibnamefont {Verstraete}}\ and\ \bibinfo {author} {\bibfnamefont {J.~I.}\ \bibnamefont {Cirac}},\ }\href {https://doi.org/10.1103/PhysRevA.70.060302} {\bibfield  {journal} {\bibinfo  {journal} {Physical Review A}\ }\textbf {\bibinfo {volume} {70}},\ \bibinfo {pages} {060302} (\bibinfo {year} {2004})}\BibitemShut {NoStop}%
\bibitem [{\citenamefont {Walshe}\ \emph {et~al.}(2025)\citenamefont {Walshe}, \citenamefont {Baragiola}, \citenamefont {Ferretti}, \citenamefont {Gefaell}, \citenamefont {Vasmer}, \citenamefont {Weil}, \citenamefont {Matsuura}, \citenamefont {Jaeken}, \citenamefont {Pantaleoni}, \citenamefont {Han}, \citenamefont {Hillmann}, \citenamefont {Menicucci}, \citenamefont {Tzitrin},\ and\ \citenamefont {Alexander}}]{walshe_linear-optical_2025}%
  \BibitemOpen
  \bibfield  {author} {\bibinfo {author} {\bibfnamefont {B.~W.}\ \bibnamefont {Walshe}}, \bibinfo {author} {\bibfnamefont {B.~Q.}\ \bibnamefont {Baragiola}}, \bibinfo {author} {\bibfnamefont {H.}~\bibnamefont {Ferretti}}, \bibinfo {author} {\bibfnamefont {J.}~\bibnamefont {Gefaell}}, \bibinfo {author} {\bibfnamefont {M.}~\bibnamefont {Vasmer}}, \bibinfo {author} {\bibfnamefont {R.}~\bibnamefont {Weil}}, \bibinfo {author} {\bibfnamefont {T.}~\bibnamefont {Matsuura}}, \bibinfo {author} {\bibfnamefont {T.}~\bibnamefont {Jaeken}}, \bibinfo {author} {\bibfnamefont {G.}~\bibnamefont {Pantaleoni}}, \bibinfo {author} {\bibfnamefont {Z.}~\bibnamefont {Han}}, \bibinfo {author} {\bibfnamefont {T.}~\bibnamefont {Hillmann}}, \bibinfo {author} {\bibfnamefont {N.~C.}\ \bibnamefont {Menicucci}}, \bibinfo {author} {\bibfnamefont {I.}~\bibnamefont {Tzitrin}},\ and\ \bibinfo {author} {\bibfnamefont {R.~N.}\ \bibnamefont {Alexander}},\ }\href {https://doi.org/10.1103/PhysRevLett.134.100602} {\bibfield  {journal} {\bibinfo
  {journal} {Physical Review Letters}\ }\textbf {\bibinfo {volume} {134}},\ \bibinfo {pages} {100602} (\bibinfo {year} {2025})}\BibitemShut {NoStop}%
\bibitem [{\citenamefont {Hillmann}\ \emph {et~al.}(2024)\citenamefont {Hillmann}, \citenamefont {Dauphinais}, \citenamefont {Tzitrin},\ and\ \citenamefont {Vasmer}}]{hillmann_single-shot_2024}%
  \BibitemOpen
  \bibfield  {author} {\bibinfo {author} {\bibfnamefont {T.}~\bibnamefont {Hillmann}}, \bibinfo {author} {\bibfnamefont {G.}~\bibnamefont {Dauphinais}}, \bibinfo {author} {\bibfnamefont {I.}~\bibnamefont {Tzitrin}},\ and\ \bibinfo {author} {\bibfnamefont {M.}~\bibnamefont {Vasmer}},\ }\href {https://doi.org/10.48550/arXiv.2410.12963} {\bibinfo {title} {Single-shot and measurement-based quantum error correction via fault complexes}} (\bibinfo {year} {2024}),\ \Eprint {https://arxiv.org/abs/2410.12963} {arXiv:2410.12963 [quant-ph]} \BibitemShut {NoStop}%
\bibitem [{\citenamefont {McEwen}\ \emph {et~al.}(2023)\citenamefont {McEwen}, \citenamefont {Bacon},\ and\ \citenamefont {Gidney}}]{mcewen_relaxing_2023}%
  \BibitemOpen
  \bibfield  {author} {\bibinfo {author} {\bibfnamefont {M.}~\bibnamefont {McEwen}}, \bibinfo {author} {\bibfnamefont {D.}~\bibnamefont {Bacon}},\ and\ \bibinfo {author} {\bibfnamefont {C.}~\bibnamefont {Gidney}},\ }\href {https://doi.org/10.22331/q-2023-11-07-1172} {\bibfield  {journal} {\bibinfo  {journal} {Quantum}\ }\textbf {\bibinfo {volume} {7}},\ \bibinfo {pages} {1172} (\bibinfo {year} {2023})}\BibitemShut {NoStop}%
\bibitem [{\citenamefont {van~de Wetering}(2020)}]{wetering_zx-calculus_2020}%
  \BibitemOpen
  \bibfield  {author} {\bibinfo {author} {\bibfnamefont {J.}~\bibnamefont {van~de Wetering}},\ }\href@noop {} {\bibinfo {title} {{ZX}-calculus for the working quantum computer scientist}} (\bibinfo {year} {2020}),\ \Eprint {https://arxiv.org/abs/2012.13966} {arXiv:2012.13966} \BibitemShut {NoStop}%
\bibitem [{\citenamefont {Gidney}(2023)}]{gidney_pair_2023}%
  \BibitemOpen
  \bibfield  {author} {\bibinfo {author} {\bibfnamefont {C.}~\bibnamefont {Gidney}},\ }\href@noop {} {\bibinfo {title} {A {Pair} {Measurement} {Surface} {Code} on {Pentagons}}} (\bibinfo {year} {2023}),\ \Eprint {https://arxiv.org/abs/2206.12780} {arXiv:2206.12780} \BibitemShut {NoStop}%
\bibitem [{\citenamefont {Bombin}\ \emph {et~al.}(2024)\citenamefont {Bombin}, \citenamefont {Litinski}, \citenamefont {Nickerson}, \citenamefont {Pastawski},\ and\ \citenamefont {Roberts}}]{bombin_unifying_2024}%
  \BibitemOpen
  \bibfield  {author} {\bibinfo {author} {\bibfnamefont {H.}~\bibnamefont {Bombin}}, \bibinfo {author} {\bibfnamefont {D.}~\bibnamefont {Litinski}}, \bibinfo {author} {\bibfnamefont {N.}~\bibnamefont {Nickerson}}, \bibinfo {author} {\bibfnamefont {F.}~\bibnamefont {Pastawski}},\ and\ \bibinfo {author} {\bibfnamefont {S.}~\bibnamefont {Roberts}},\ }\href {https://doi.org/10.22331/q-2024-06-18-1379} {\bibfield  {journal} {\bibinfo  {journal} {Quantum}\ }\textbf {\bibinfo {volume} {8}},\ \bibinfo {pages} {1379} (\bibinfo {year} {2024})}\BibitemShut {NoStop}%
\bibitem [{\citenamefont {Tomita}\ and\ \citenamefont {Svore}(2014)}]{tomita_low-distance_2014}%
  \BibitemOpen
  \bibfield  {author} {\bibinfo {author} {\bibfnamefont {Y.}~\bibnamefont {Tomita}}\ and\ \bibinfo {author} {\bibfnamefont {K.~M.}\ \bibnamefont {Svore}},\ }\href {https://doi.org/10.1103/PhysRevA.90.062320} {\bibfield  {journal} {\bibinfo  {journal} {Physical Review A}\ }\textbf {\bibinfo {volume} {90}},\ \bibinfo {pages} {062320} (\bibinfo {year} {2014})}\BibitemShut {NoStop}%
\bibitem [{\citenamefont {Gidney}(2021)}]{gidney_stim_2021}%
  \BibitemOpen
  \bibfield  {author} {\bibinfo {author} {\bibfnamefont {C.}~\bibnamefont {Gidney}},\ }\href {https://doi.org/10.22331/q-2021-07-06-497} {\bibfield  {journal} {\bibinfo  {journal} {Quantum}\ }\textbf {\bibinfo {volume} {5}},\ \bibinfo {pages} {497} (\bibinfo {year} {2021})}\BibitemShut {NoStop}%
\bibitem [{\citenamefont {Higgott}\ and\ \citenamefont {Gidney}(2025)}]{higgott_sparse_2025}%
  \BibitemOpen
  \bibfield  {author} {\bibinfo {author} {\bibfnamefont {O.}~\bibnamefont {Higgott}}\ and\ \bibinfo {author} {\bibfnamefont {C.}~\bibnamefont {Gidney}},\ }\href {https://doi.org/10.22331/q-2025-01-20-1600} {\bibfield  {journal} {\bibinfo  {journal} {Quantum}\ }\textbf {\bibinfo {volume} {9}},\ \bibinfo {pages} {1600} (\bibinfo {year} {2025})}\BibitemShut {NoStop}%
\end{thebibliography}%
